\documentstyle[amssymb,psfig]{mn2e}

\title[The NGC 5253 star cluster system. I.]{The NGC 5253 star cluster
  system. I. Standard modelling and infrared-excess sources}

\author[Richard de Grijs et al.]{Richard de
  Grijs,$^{1,2,3}$\thanks{E-mail: grijs@pku.edu.cn} Peter
  Anders,$^{1}$ Erik Zackrisson$^4$ and G\"oran \"Ostlin$^4$\\
$^1$ Kavli Institute for Astronomy and Astrophysics, Peking
University, Yi He Yuan Lu 5, Hai Dian District, Beijing 100871,
China\\
$^2$ Department of Astronomy and Space Science, Kyung Hee University,
Yongin-shi, Kyungki-do 449-701, Republic of Korea\\
$^3$ 2012 Selby Fellow, Australian Academy of Science\\
$^4$ Department of Astronomy, Stockholm University, Oscar Klein
Centre, AlbaNova, Stockholm SE-106 91, Sweden
}

\date{Received date; accepted date}
\pubyear{2012}

\begin{document}
\maketitle

\begin{abstract}
Using high-resolution {\sl Hubble Space Telescope} data, we reexamine
the fundamental properties (ages, masses and extinction values) of the
rich star cluster population in the dwarf starburst galaxy NGC
5253. The gain in resolution compared to previous studies is of order
a factor of two in both spatial dimensions, while our accessible
wavelength range transcends previous studies by incorporation of both
near-ultraviolet and near-infrared (IR) passbands. We apply spectral
synthesis treatments based on two different simple stellar population
model suites to our set of medium-, broad-band and H$\alpha$ images to
gain an improved physical understanding of the IR-excess flux found
for a subset of young clusters (30 of 149). With the caveat that our
models are based on fully sampled stellar mass functions, the NGC 5253
cluster population is dominated by a significant number of relatively
low-mass ($M_{\rm cl} \la \mbox{ a few} 10^4$ M$_\odot$) objects with
ages ranging from a few $\times 10^6$ to a few $\times 10^7$ yr, which
is in excellent agreement with the starburst age of the host
galaxy. The IR-excess clusters are almost all found in this young age
range and have masses of up to a few $\times 10^4$ M$_\odot$. The IR
excess in the relatively low-mass NGC 5253 clusters is most likely
caused by a combination of stochastic sampling effects and colour
variations due to the presence of either luminous red or
pre-main-sequence stars. We also find a small number of
intermediate-age ($\sim 1$ Gyr-old), $\sim10^5$ M$_\odot$ clusters, as
well as up to a dozen massive, $\sim 10$ Gyr-old globular
clusters. Their presence supports the notion that NGC 5253 is a very
active galaxy that has undergone multiple episodes of star cluster
formation.
\end{abstract}


\section{Introduction}
\label{intro.sec}

\subsection{Choice of target}

NGC 5253, a metal-poor blue compact dwarf galaxy in the Centaurus A
group (Karachentsev et al. 2007), hosts an extremely young
starburst. At the fortuitously close distance of 3.13 Mpc (Davidge
2007; 1 arcsec corresponds to 16 pc), the galaxy has been the target
of a large number of observational programmes spanning a range of
wavelengths, from X-rays to the radio domain (e.g., Rieke, Lebofsky \&
Walker 1988; Caldwell \& Phillips 1989; Martin \& Kennicutt 1995;
Freedman et al. 2001; Cresci, Vanzi \& Sauvage 2005;
Mart\'{\i}n-Hern\'andez, Schaerer \& Sauvage 2005; L\'opez-S\'anchez
et al. 2007, 2011).

The metallicity of NGC 5253, approximately 0.2--0.4 Z$_\odot$
(Kobulnicky \& Skillman 1995; Kobulnicky et al. 1999;
L\'opez-S\'anchez et al. 2011), is relatively constant across the
galaxy, except for a few areas exhibiting higher-than-average nitrogen
abundances (Walsh \& Roy 1989; Kobulnicky et al. 1997;
L\'opez-S\'anchez et al. 2007; L\'opez-S\'anchez \& Esteban 2010a,b;
Monreal-Ibero et al. 2010) and a possible slight helium enrichment
(Campbell, Terlevich \& Melnick 1986; L\'opez-S\'anchez et
al. 2007). This low metallicity, combined with the galaxy's relatively
small dimensions ($\sim 5 \times 2$ arcmin$^2 \simeq 4.6 \times 1.8$
kpc$^2$ major/minor axis extents) and low mass ($\sim 2$--$5 \times
10^9$ M$_\odot$; Bottinelli et al. 1972; Turner, Beck \& Hurt 1997),
render its starburst component an excellent analogue to galaxies in
their early formation phases, while its proximity makes it ideal for
studies of the many bright, young stellar clusters (YSCs; e.g.,
Gonzalez-Riestra, Rego \& Zamorano 1987; Caldwell \& Phillips 1989;
Calzetti et al. 1997; Harris et al. 2004; Vanzi \& Sauvage 2004;
Cresci et al. 2005; Mart\'{\i}n-Hern\'andez et al. 2005; Harbeck,
Gallagher \& Crnojevi\'c 2012).

In fact, based on ultraviolet (UV) {\sl Hubble Space Telescope
  (HST)}/Space Telescope Imaging Spectrometer (STIS) spectroscopy of
the YSCs and diffuse background light in the main body of the galaxy,
Tremonti et al. (2001) suggested that star clusters may have been
forming continuously, and subsequently dissolve on $\sim$ 10--20 Myr
timescales (based on arguments related to tidal effects worked out in
detail in Kim, Morris \& Lee 1999; a process since coined cluster
`infant mortality'), thus dispersing their stars into the field star
population.

\subsection{Starburst characteristics}

The galaxy's H$\alpha$ morphology (Martin 1998; Calzetti et al. 2004;
Meurer et al. 2006) shows that the inner core is currently undergoing
intense star formation and is, in fact, one of the youngest starbursts
known (van den Bergh 1980; Moorwood \& Glass 1982; Rieke et al. 1988;
Caldwell \& Phillips 1989; Calzetti et al. 1997; McQuinn et
al. 2010a,b). The active starburst is confined to the inner 60 pc
(Calzetti et al. 1997, 1999; Tremonti et al. 2001); age estimates for
the starburst duration in this region are $< 1 \times 10^7$ yr, and
more likely a few $\times 10^6$ yr, assuming a burst-like formation
history for the diffuse stellar population in the disc (Walsh \& Roy
1989; Rieke et al. 1988; Tremonti et al. 2001). These estimates are
based on weak [Fe {\sc ii}] emission, strong Br$\gamma$ emission, a
lack of significant (non-thermal) synchrotron emission from supernova
remnants (Beck et al. 1996), the presence of Wolf--Rayet stars
(Campbell et al. 1986; Kobulnicky et al. 1997; Schaerer et al. 1997;
L\'opez-S\'anchez et al. 2007; L\'opez-S\'anchez \& Esteban 2010a,b;
Monreal-Ibero et al. 2010) and weak $2 \mu$m emission, implying only a
small contribution of giants and supergiants, which thus provides a
rough upper age limit.

In two very detailed studies of the galaxy's central starburst region,
Calzetti et al. (1997, 1999) showed that the UV emission is dominated
by a 3--4 Myr old, relatively small star cluster, NGC 5253-4 (Meurer
et al. 1995). On the other hand, a dust-enshrouded ($A_V = 9$--35 mag)
very young central cluster (2--2.7 Myr old), NGC 5253-5, is
responsible for most of the nuclear region's ionization (Tremonti et
al. 2001; Cresci et al. 2005). At radio wavelengths, Turner et
al. (1998, 2000) derive ionization potentials driven by large numbers
(200--1000) of O-type stars, which they interpret to imply that very
large clusters are the preferred mode of star formation in the central
regions of the galaxy.

\subsection{Current context}

It is thus clear that the galaxy's star and star cluster formation
history is highly complex. Despite the plethora of existing, detailed
studies, many of the salient details of the galaxy's evolution remain
open-ended. This realization is perhaps mostly driven by the object's
proximity, which allows us to probe the star-forming environment in
significantly greater detail than in more distant starbursts. On the
other hand, NGC 5253's proximity also offers us the key advantage of
it having been observed repeatedly across the full observable 
wavelength range.

Nevertheless, where the stellar component has been subject to detailed
scrutiny, most studies have either focussed on careful, mostly
spectroscopic studies of a few objects at a time, or systematic
exploration of the galaxy's full extent but based on relatively simple
approaches. This is particularly so for its star cluster
population. Previous statistical studies have relied on
straightforward application of simple stellar population (SSP)
analysis based on optical broad-band data (e.g., Harris et al. 2004;
Cresci et al. 2005; Harbeck et al. 2012). However, in recent years,
theory and computational approaches have made significant advances.

In this paper, we go significantly beyond previous studies of the NGC
5253 star cluster population in a number of complementary ways. First,
although we still use primarily (medium- and) broad-band imaging
observations, we base our results on the highest-achievable spatial
resolution from the near-UV (NUV) to the near-infrared (NIR)
regimes. The gain in resolution compared to previous studies is of
order a factor of two in both spatial dimensions, while our accessible
wavelength range transcends previous studies by incorporation of the
crucial NUV and NIR passbands (which will allow us to at least
partially break the age--metallicity and age--extinction
degeneracies). Where relevant, we also refer to H$\alpha$
emission-line results to further reduce the uncertainties.

Second, we focus specifically on one aspect of extragalactic star
cluster populations where we take full advantage of the observational
gains, i.e., the importance and origin of a red or IR excess in the
spectral-energy distributions (SEDs) of a relatively low-mass cluster
population such as that in NGC 5253. To address this aspect, we do not
need to have access to a statistically complete sample of star
clusters down to the lowest detection limits enabled by the
observations' worst signal-to-noise ratio (S/N).

This paper is organized as follows. In Section \ref{observations.sec}
we discuss the observational data which form the basis of our
analysis, as well as our data reduction approach. The analysis methods
adopted are outlined in Section \ref{analysis.sec}, while we discuss
the robustness of the results and perform a comparison in Section
\ref{robust.sec}. We offer a preliminary discussion of the effects of
stochasticity in the stellar mass function separately, in Section
\ref{stoch.sec}, and provide an overview of the physical state of the
NGC 5253 cluster population in Section \ref{overall.sec}. Finally, we
summarize and conclude the paper in Section \ref{summary.sec}.

This is the first article in a series of three. In Paper II (P. Anders
et al., in prep.), we will present a detailed theoretical
investigation of the effects of stochastic sampling on cluster age and
mass determinations based on adopting fully sampled stellar mass
functions using our own SSP models and analysis approach (cf. Section
\ref{analysis.sec}). In Paper III (R. de Grijs et al., in prep.), we
will apply these insights to the data set analysed here, to explore in
detail to which extent our results are affected by biases owing to
stochastic sampling of the clusters' mass functions. The present paper
is meant to establish the baseline for this body of work.

\section{Hubble Space Telescope observations and data reduction}
\label{observations.sec}

\subsection{Data and basic reduction}

We mined the {\sl HST} Data Archive for available medium-/broad-band
and H$\alpha$ observations of NGC 5253 obtained with its main optical
and NIR cameras, i.e., the Advanced Camera for Surveys (ACS), the
Wide-Field and Planetary Camera-2 (WFPC2) and the Near-Infrared Camera
and Multi-Object Spectrometer (NICMOS). Since we had selected our
target galaxy based on the availability of a large number of
high-resolution {\sl HST} data sets spanning the longest possible
wavelength range, we could make a careful selection of the best
possible data sets to reach our science goals. Table \ref{data.tab}
includes details of the final set of observations used, which were
selected to achieve the best possible combination of spatial
resolution (aimed at reducing contamination by neighbouring sources),
S/N (for which we initially used exposure time as a proxy) and
extensive, continuous wavelength coverage.

\begin{table*}
\caption[ ]{\label{data.tab}{\sl HST} observations used for the analysis in this paper.}
\begin{center}
\begin{tabular}{ccclrrlc}
\hline
\hline
Filter & Camera & Chip & Exposures (s) & \multicolumn{1}{c}{Total exp.} & Proposal & \multicolumn{1}{c}{PI} & \multicolumn{1}{c}{Photometric corr.} \\
       &        &      &             & time (s) & \multicolumn{1}{c}{ID} & & \multicolumn{1}{c}{factor (unit)}\\
\hline
F330W & ACS    & HRC  & $3 \times 449$               & 1347\ \ \ \ \ & 10609 & Vacca    & 1.0\ \ \ ($\sigma_{\rm HRC}$) \\
F435W & ACS    & HRC  & $4 \times 150$               &  600\ \ \ \ \ & 10609 & Vacca    & 1.1\ \ \ ($\sigma_{\rm HRC}$) \\
F547M & WFPC2  & WF   & $2 \times 200, 2 \times 600$ & 1600\ \ \ \ \ &  6524 & Calzetti & 1.6\ \ \  ($\sigma_{\rm WFC}$) \\
F550M & ACS    & HRC  & $4 \times 200$               &  800\ \ \ \ \ & 10609 & Vacca    & 1.0\ \ \  ($\sigma_{\rm HRC}$) \\
F555W & ACS    & WFC  & $4 \times 600$               & 2400\ \ \ \ \ & 10765 & Zezas    & 1.0\ \ \  ($\sigma_{\rm WFC}$) \\
F606W & WFPC2  & PC   & $500$                        &  500\ \ \ \ \ &  5479 & Malkan   & 2.0\ \ \  ($\sigma_{\rm HRC}$) \\
F658N & ACS    & HRC  & $240$                        &  240\ \ \ \ \ & 10609 & Vacca    & 1.0\ \ \  ($\sigma_{\rm HRC}$) \\
F814W & ACS    & HRC  & $4 \times 92$                &  368\ \ \ \ \ & 10609 & Vacca    & 1.25 ($\sigma_{\rm HRC}$) \\
F110W & NICMOS & NIC2 & $4 \times 23.97$             &   95.86 &  7219 & Scoville       & 1.2\ \ \  ($\sigma_{\rm WFC}$) \\
F160W & NICMOS & NIC2 & $4 \times 23.97$             &   95.86 &  7219 & Scoville       & 1.1\ \ \  ($\sigma_{\rm WFC}$) \\
F222M & NICMOS & NIC2 & $4 \times 63.95$             &  255.80 &  7219 & Scoville       & 1.15 ($\sigma_{\rm WFC}$) \\
\hline
\end{tabular}
\end{center}
\flushleft
Note: The three-digit numbers in the filter names represent their
effective wavelengths in nanometres; the final digits indicate the
filter type: N = narrow, M = medium, W = wide.
\end{table*}

\subsection{Object selection}
\label{objects.sec}

We used the ACS/HRC data in the F330W, F435W, F550M and F814W filters
as basis for the final orientation of all observations. This implies
that we rotated, scaled and aligned all other images to the position,
scale and orientation of the ACS/HRC data using the {\sc iraf/stsdas}
tasks\footnote{The Image Reduction and Analysis Facility ({\sc iraf})
  is distributed by the National Optical Astronomy Observatories,
  which is operated by the Association of Universities for Research in
  Astronomy, Inc., under cooperative agreement with the U.S. National
  Science Foundation. {\sc stsdas}, the Space Telescope Science Data
  Analysis System, contains tasks complementary to the existing {\sc
    iraf} tasks. We used Version 3.6 (November 2006) for the data
  reduction performed in this paper.} {\sc rotate}, {\sc magnify}, and
{\sc imalign}, respectively, using a selection of conspicuous sources
common to each frame as guides. We checked that our image manipulation
did not affect the reliability of the integrated flux measurements of
our sample of candidate star clusters. The final image size of the set
of aligned images is $1169 \times 1138$ ACS/HRC pixels,\footnote{The
  ACS/HRC pixels are rectangular and subtend $0.028 \times 0.025$
  arcsec$^2$ each.} corresponding to $32.73 \times 28.45$ arcsec$^2$
($\sim 500 \times 430$ pc$^2$).

The standard deviations ($\sigma_{\rm sky}$) of the number of counts
in seemingly empty sections in all images were established to
ascertain a `sky' background count for each filter. Multiples of this
background count were used as thresholds above which the numbers of
`real' sources in both the F550M and F814W ACS/HRC filters were
calculated, using the {\sc idl}\footnote{The Interactive Data Language
  ({\sc idl}) is licensed by Research Systems Inc., of Boulder, CO,
  USA.} {\sc find} task. We did not force our detection routine to
constrain the sources' roundness or sharpness, in order to be as
inclusive as possible. The most suitable thresholds for source
inclusion, at $4 \sigma_{\rm sky}$ (see below for justification), were
0.7 and 1.5 counts s$^{-1}$ (270 and 248 source detections) for the
F550M and F814W filters, respectively. In all passbands, the number of
detections initially decreases rapidly with increasing threshold
value. This is an indication that our `source' detections are
noise-dominated in the low-threshold regime. Where the rapid decline
slows down to a more moderate rate, our detections become dominated by
`real' objects (either stars, clusters or cluster candidates, or real
intensity variations in the background field: for an example, see
fig. 3 of Barker et al. 2008). In the remainder of this paper we will
only consider the objects in the `source-dominated' domain, with
fluxes in excess of the relevant $4 \sigma_{\rm sky}$ threshold.

We subsequently employed a cross-identification procedure to determine
how many sources coincided with intensity peaks within 1.4 pixels of
each other in the F550M$\otimes$F814W HRC filter combination (i.e.,
allowing for 1-pixel mismatches in both spatial directions). In our
next step, we applied a Gauss-fitting routine in {\sc idl} to each
candidate cluster. We also determined the best-fitting Gaussian
profiles for some 10--12 isolated, star-like sources in all filters
(where the number of objects available for this purpose depended on
the S/N of the observations and the typical stellar spectra) and thus
obtained observational stellar Gaussian $\sigma_{\rm G}$ (size) values
for all filters. The Gaussian stellar widths in the F550M HRC band
(the filter we used for our size selection) were found to be best
represented by a Gaussian profile of $\sigma_{\rm G} = 1.1$
pixels. Any candidate cluster with $\sigma_{\rm G,F550M} \le 1.1$
pixels was considered most likely to be a star (or -- if much smaller
than 1 pixel -- an artefact either of the detector or caused by cosmic
rays) and was consequently discarded. We applied a conservative
clipping criterion so as not to reject some marginally extended
sources. These steps resulted in rejection of 88 objects from the
original F550M source list and led to a working sample containing 182
extended cluster candidates.

\begin{figure}
\psfig{figure=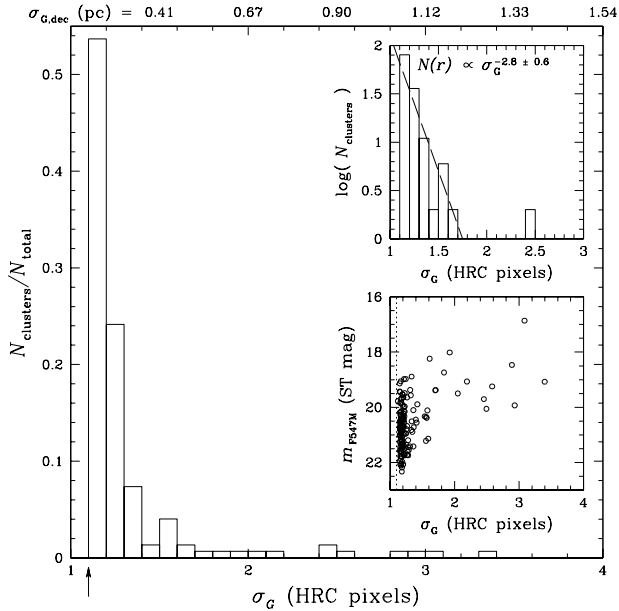,width=\columnwidth}
\caption{\label{sizes.fig}Gaussian size distribution of the extended
  star cluster candidates in our sample (as fraction of the total
  sample). The arrow indicates our minimum size cut-off, corresponding
  to the $\sigma_{\rm G}$ of point sources. The top axis shows the
  linear cluster sizes after deconvolution of the observed
  $\sigma_{\rm G}$ and the intrinsic size of the PSF. The insets show
  the size distribution {\it (top)} in logarithmic units, combined
  with the best-fitting power-law function (dashed line, defined by
  the proportionality given) and {\it (bottom)} as a function of F547M
  magnitude (see Section \ref{sizes.sec} for discussion). The vertical
  dotted line indicates the ACS/HRC PSF size.}
\end{figure}

Their size distribution is shown in Fig. \ref{sizes.fig}, where we
also show the linear cluster sizes after deconvolution of the observed
$\sigma_{\rm G}$ and the intrinsic size of the point-spread function
(PSF). Our (candidate) cluster sample is characterized by a peak at
sizes close to that of the PSF, while the most extended cluster
candidate has $\sigma_{\rm G} = 3.4$ pixels (equivalent to 1.36 pc
after PSF deconvolution). Our sample is insensitive to stellar
associations because of the selection criteria adopted. We will
discuss the implications of the cluster size distribution (as well as
the insets in Fig. \ref{sizes.fig}) in Section \ref{sizes.sec}. In a
final step, we confirmed that our procedure is robust with respect to
misidentifications of genuine clusters as point sources (stars): only
a few ($\la 5$) per cent of the objects that were rejected based on
the Gaussian size estimates were brighter than the tip of the red
giant branch (TRGB) magnitude at the distance of NGC 5253, where we
used the most recent calibration based on observations of $\omega$
Centauri, $M_I^{\rm TRGB} = -4.04 \pm 0.12$ mag (Bellazzini et
al. 2001, 2004).

We note that although realistic cluster luminosity profiles may well
deviate from the simple Gaussian profile adopted here, its consistent
and systematic application to our observations allows us to
differentiate accurately between objects of different sizes,
irrespective of their true profiles, provided that the intrinsic
profiles of the individual objects do not differ too much from object
to object (although one should realize that stochastic sampling
effects may invalidate this assumption; see below). In
Fig. \ref{sources.fig} we show the distribution of our candidate
cluster sample across the face of the NGC 5253 common field in the
ACS/HRC F550M filter.\footnote{For a movie including all
  high-resolution images and object identifications, see
  http://kiaa.pku.edu.cn/$\sim$grijs/ngc5253.mov.} The main
differences between our current sample selection and that of Cresci et
al. (2005) -- the most recent detailed statistical study of the NGC
5253 cluster system -- are owing to the higher-resolution images we
used as the basis for our object identification, thus reducing
crowding effects. As a result, we find slightly fewer cluster
candidates in the galaxy's main body; some of the sources identified
by Cresci et al. (2005) as `clusters' are unresolved (and likely
stars) in our images (with $\sigma_{\rm G} \le 2.65$ pc, corresponding
to a PSF-deconvolved size of $\sigma_{\rm G,dec} \le 2.4$ pc).

\begin{figure*}
\hspace*{1cm}
\psfig{figure=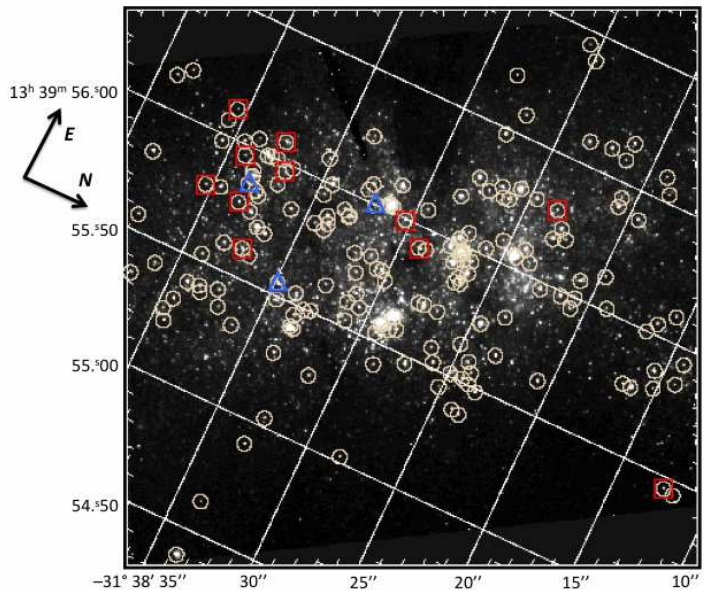,width=11cm}
\caption{\label{sources.fig}NGC 5253 field of view in common of all
  observations and source selection discussed in this paper, overlaid
  on the ACS/HRC F550M image. Red squares: Age $t \simeq 10^{10}$ yr;
  blue triangles: $t \simeq 10^9$ yr (for a discussion of the
  clusters' spatial distribution as a function of age, see
  Section \ref{agedist.sec}).}
\end{figure*}

\subsection{Photometry}
\label{photom.sec}

Our custom-written {\sc idl} aperture photometry task uses source
radii and sky annuli {\it individualized} for each cluster
candidate. The adopted radii and annuli were based on Gaussian stellar
size measurements ($\sigma_{\rm G}$) obtained in the F550M filter. We
applied a correction factor as a function of filter/instrument to
correct for different intrinsic PSF sizes (see the final column in
Table \ref{data.tab}), based on determination of the best-fitting
Gaussian profiles for the 10--12 isolated, star-like sources in all
filters we used for our size cuts in Section \ref{objects.sec}. We
used the F550M HRC and F555W ACS/Wide Field Camera (WFC) images as our
basis for the determination of source and sky annuli because of their
high resolution and good S/N: in images with HRC (or similar, e.g.,
WFPC2/PC) resolution, we used a source aperture radius of $4\times
\mbox{[} \sigma_{\rm F550M} \times$ HRC correction factor], and 4 and
$6\times \mbox{[} \sigma_{\rm F550M} \times$ HRC correction factor]
for the inner and outer sky annuli (in pixels), respectively. For the
other images, characterized by poorer resolution, the relevant numbers
were 3, 3.5 and $5 \times \mbox{[}\sigma_{\rm F555W} \times$ WFC
  correction factor], respectively, again expressed in pixel units
(the WFC chips are characterized by a pixel size of 0.049 arcsec). The
units in the final column of Table \ref{data.tab} corresponding to
$\sigma_{\rm HRC}$ and $\sigma_{\rm WFC}$ refer to $\sigma_{\rm
  F550M}$ and $\sigma_{\rm F555W}$, respectively.

This choice of apertures and their scaling with the objects' measured
sizes was based on extensive tests in which we inspected the stellar
radial profiles, to identify where the object profiles generally
disappear into the background noise. Specifically, we carefully
checked the profiles of all candidate clusters, for all filters and
detectors, to verify that our source radii were chosen conservatively
and sufficiently far out so as not to exclude any genuine source flux
with respect to the general background level (i.e., well in excess of
the radii where the radial profiles disappear into the background
noise). We also checked that the background annuli were chosen
appropriately and not dominated by neighbouring sources. Finally, we
checked that the background level was generally flat (within the
photometric uncertainties and the statistical noise) so that small
differences among the radii used to define the background annuli as a
function of filter would not cause artificial offsets in flux, and
hence lead to colour differences.

\begin{figure}
\psfig{figure=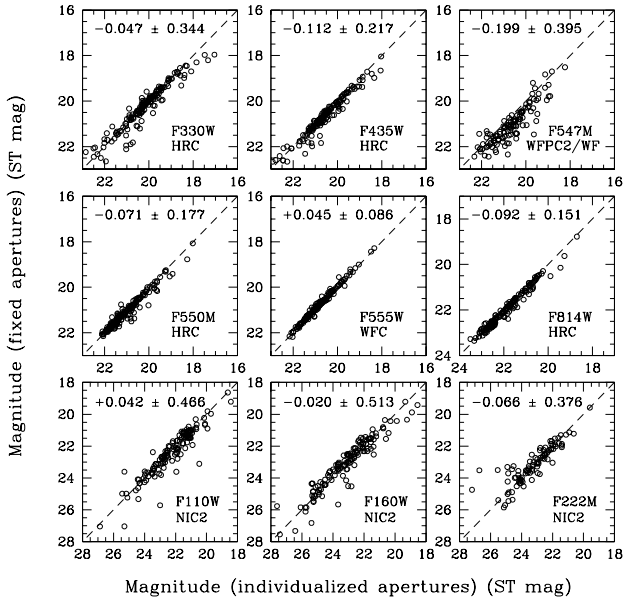,width=\columnwidth}
\caption{\label{comparemags.fig}Comparison of aperture photometry
  based on our individualized apertures and that from the use of fixed
  apertures. For this example, we employed source apertures of 8
  pixels (in the detector's native system) and background annuli
  between 8 and 12 pixels for our ACS-based fixed-aperture
  photometry. For the F547M WFPC2/WF and the NICMOS images, we used
  equivalent radii of 12, 12 and 15 pixels, again in the respective
  detectors' native systems. The photometric data points in the
  different panels only include those objects for which the
  individualized apertures were smaller than or equal to the sizes of
  our fixed apertures, so that a direct one-to-one comparison is
  possible. The mean differences in the sense $(m_{\rm indiv.\; ap.} -
  m_{\rm fixed\; ap.})$, as well as the standard deviations of these
  difference distributions are included in each of the panels for the
  relevant instrument/filter combination (also indicated). The dashed
  lines indicate the loci of equality.}
\end{figure}

Fig. \ref{comparemags.fig} shows a typical example of the level of
photometric accuracy attainable with individualized apertures compared
to the use of fixed apertures. For this example, we employed source
apertures of 8 pixels (in the detector's native system) and background
annuli between 8 and 12 pixels for our ACS-based fixed-aperture
photometry. For the F547M WFPC2/WF and the NICMOS images, we used
equivalent radii of 12, 12 and 15 pixels, again in the respective
detectors' native systems. The photometric data points in the
different panels only include those objects for which the
individualized apertures were smaller than or equal to the sizes of
our fixed apertures, so that a direct one-to-one comparison is
possible. The mean differences in the sense $(m_{\rm indiv. \;ap.} -
m_{\rm fixed\; ap.})$, as well as the standard deviations of these
difference distributions are included in each of the panels for the
relevant instrument/filter combination. It is clear that the
correspondence between our individualized aperture photometry and that
obtained from using fixed apertures (chosen so that most sources would
have most of their flux recorded) is tight. The scatter increases for
the lower-resolution and lower-S/N images, although the mean
difference remains small, particularly within the associated
uncertainties.

For the size cut-offs ($1.1 \sigma_{\rm G}$) in the HRC F550M and the
WFC F555W filters, the annuli adopted in this study correspond to
deconvolved linear sizes of 2.35, 2.35 and 3.50 pc, and 2.45, 3.85 and
4.10 pc, respectively. The resulting instrumental magnitudes of the
182 `clusters' for which our tasks could obtain reliable photometry
(with photometric uncertainties of $< 0.3$ mag in at least 7 filters)
were adjusted for their zero-point offsets (determined from the
PHOTFLAM and PHOTZPT header keywords of the {\sl HST} observations,
thus resulting in photometry in the ST mag system). This sample forms
the basis of our analysis in Sections \ref{robust.sec} and
\ref{stoch.sec}. We will make the full photometric data tables of all
182 cluster candidates available in Paper III, upon completion of our
detailed investigation of the effects of stochasticity in the
clusters' stellar mass functions.

Note that because of the extended nature of our objects (which was,
after all, a key selection criterion), we could not apply PSF
photometry or simple aperture corrections (but see, e.g., Harris et
al. 2004; Annibali et al. 2011); the latter would require uncertain
assumptions about, e.g., the underlying source profile (cf. Anders,
Gieles \& de Grijs 2006) as well as the degree of mass segregation
within the clusters. The rationale for such an approach is to attempt
to obtain the full flux of a given object. Particularly for the
lower-luminosity, lower-mass objects, the effects of stochastic
sampling would likely introduce additional uncertainties if adopting a
generic underlying radial profile. On the other hand, adopting an
aperture correction to large radii based on the profile and colours in
a restricted region relatively close to a cluster's core may give rise
to additional colour- and age-dependent uncertainties in the derived
total fluxes owing to the effects of stellar mass
segregation. Instead, we opted to use a hybrid method: our aperture
sizes were variable, i.e., based on the measured cluster sizes and
designed such that our output photometry included the full flux
contribution from all objects.\footnote{In a sense, whether or not one
  should (or even could) use aperture corrections to recover the
  fluxes of extended objects is a philosophical issue. In this paper,
  we decided to adopt the lessons learnt from the results of Anders et
  al. (2006) and choose source apertures based on the objects' size
  measurements.}

In addition, our integrated SEDs are dominated by the
highest-luminosity (inner) regions of the clusters (we checked that
any contribution from potentially missed flux in the outer regions is
negligible), so that the resulting parameters do not strongly depend
on our aperture photometry approach (see also Section
\ref{halpha.sec}). Finally, the majority of our sample clusters are
sufficiently isolated that our approach is not severely affected by
the effects of crowding; this is supported by inspection of the growth
curves of the dozen sample clusters in the most crowded
regions. Ultimately, the main advantage of our hybrid approach is that
it is simple and requires few, if any, assumptions that may introduce
additional uncertainties.

\begin{figure}
\psfig{figure=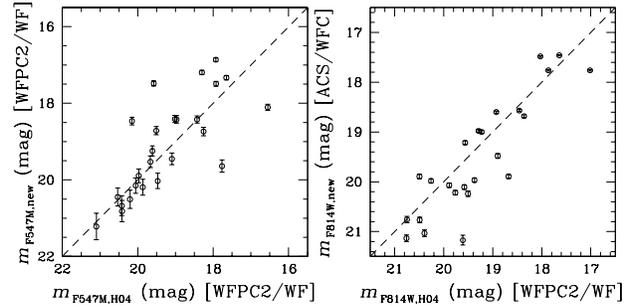,width=\columnwidth}
\vspace{-4cm}
\caption{\label{harris.fig}Comparison of our photometry using a hybrid
  approach (indicated by the subscript `new') with that of Harris et
  al. (2004), based on aperture corrections appropriate for point
  sources (referred to as H04), both in the ST mag system.}
\end{figure}

Fig. \ref{harris.fig} compares the resulting photometry with that
published by Harris et al. (2004). The latter authors published ST
magnitudes for 33 objects they identified as clusters, 25 of which are
also included in our cluster sample. We note that the overall
agreement between both data sets is reasonable, although in half a
dozen cases there are significant offsets. The latter occur
particularly for the brighter objects in the F547M filter.

Harris et al. (2004) discuss the best approach to obtaining reliable
photometry in the crowded environment of NGC 5253, which is
additionally affected by significant background variations. They
conclude that PSF-like photometry is inappropriate under these
conditions, and that aperture corrections depend sensitively on the
underlying luminosity profile adopted. They settled on achieving a
balance between the systematic errors imposed by small apertures and
the random errors inherent to adopting large apertures by selecting an
intermediate aperture size of 5 pixels (radius). They then attempt to
correct for the clusters' extended profiles by adopting a 0.1 mag,
5-pixel aperture correction appropriate for point sources, but they
caution that it is likely that there remains a systematic error in the
photometry because of the clusters' non-spherical, extended
morphologies. In this context, the offsets in the F547M photometry
between our determinations and those of Harris et al. (2004) can be
understood fairly easily. For the brigher, extended sources in
particular, the 5-pixel apertures they adopted are too small to
encompass the full source fluxes, thus leading to underestimated
luminosities (fainter magnitudes) on average. This is evident from
inspection of Fig. \ref{harris.fig}, particularly for objects with
$m_{\rm F5547M} \la 20$ mag (in both sets of photometric
measurements). In addition, two bright objects ($m_{\rm F5547M,H04}
\la 18$ mag) have brighter magnitudes in the data set of Harris et
al. (2004) than determined by us. These objects are found in the most
crowded regions of the galaxy, where their extended profiles overlap
with neighbouring objects. On balance, we prefer to use our own newly
determined photometry, because this attempts to include and correct
for the effects of non-sphericity and spatial extent. We also point
out that a comparison of our photometry based on different apertures
(shown in Fig. \ref{comparemags.fig}) indicates that our measurements
are internally consistent.

The scatter in the F814W photometry is somewhat larger than that for
the F547M data, which is most likely caused by the difference in
resolution between the cameras used in the presence of fairly crowded
fields. Lower-resolution observations, such as the WFPC2 data of
Harris et al. (2004), are more significantly affected by close
neighbours (`blending') than the higher-resolution ACS data we use in
this paper. In addition, there is a small systematic offset towards
brighter magnitudes for our photometry compared to the Harris et
al. (2004) data set, which is most likely caused by their
underestimated aperture corrections for point sources.

In Section \ref{analysis.sec} we analyse our cluster photometry based
on the objects' colours. In essence, we use a poor man's approach to
cluster spectroscopy based on the clusters' broad-band SEDs.

\section{SED analysis}
\label{analysis.sec}

We base our SED analysis on two of the current-best, most up-to-date
SSP model suites available, specifically the Yggdrasil models (Section
\ref{redxs.sec}) and the {\sc galev} models (Section
\ref{galev.sec}). Both sets of models are based on different, modern
isochrones and independently developed spectral synthesis codes, which
have all been extensively validated. Although there are currently at
least a handful of suitable SSP model suites to choose from, the two
model sets we adopted are most appropriate in the context of the
present paper, since a number of our co-authors are actively involved
in their development. This offers the key advantage that we fully
understand the basic underlying assumptions and essential physical
inputs, which we summarize in detail in this section.

\subsection{General model assumptions}

For a one-to-one comparison between the observed and model SEDs, we
assumed that star clusters are coeval (single-burst) stellar
populations characterized by a Kroupa (2001) IMF. This IMF is
characterized by a lower cut-off mass of 0.15 M$_\odot$, whereas the
upper-mass cut-off ranges between $\sim 66$ and $\sim 76$ M$_\odot$,
and is determined by the mass coverage of the Padova isochrones (i.e.,
Marigo et al. 2008) for a given metallicity, e.g., the upper mass
cut-off for $Z=0.004 \; (0.008)$ is 68.07 (70.35) M$_\odot$. The
theoretical stellar libraries are from Lejeune et al. (1997, 1998) for
a broad range of metallicities.

The current versions of both sets of models include the important
thermally pulsing asymptotic-giant-branch (TP-AGB) evolutionary
phase. At ages from $\sim 100$ Myr to $\sim 1$ Gyr, TP-AGB stars
account for 25 to 40 per cent of the bolometric light, and for 50 to
60 per cent of the $K$-band emission of SSPs (see Charlot 1996; Schulz
et al. 2002). The effect of including the TP-AGB phase results in
redder colours for SSPs with ages between $\sim 10^8$ and $10^9$ yr,
with the strongest effect (up to $\ga 1$ mag) in $(V - K)$ for solar
metallicity, and in $(V - I)$ for $Z \ge 0.5$
Z$_\odot$. Shorter-wavelength colours and lower-metallicity SSPs are
less affected.

The models predict the contributions to the combined SED of both the
stars and the ambient, photo-ionized gas owing to the presence of
young massive stars. The photo-ionized gas contributes both to the
continuum and to specific emission lines, thus directly affecting the
integrated broad-band fluxes of YSCs. Anders \& Fritze-v. Alvensleben
(2003) and Adamo et al. (2010b) showed that nebular emission
non-negligibly affects cluster SEDs during the first 10--15 Myr of
their evolution. Although this contribution reduces after 6 Myr, at
least for shorter (bluer) wavelengths, it remains significant in the
NIR regime up to $\sim 20$ Myr (depending on metallicity).

\subsection{Yggdrasil models and IR-excess analysis}
\label{redxs.sec}

{\it Model-specific details.} The primary approach we adopted for our
analysis in this paper uses the most up-to-date Yggdrasil spectral
synthesis models (Zackrisson et al. 2011). The latter are based on
Padova-AGB SSPs from the Starburst99 spectral synthesis code
(Leitherer et al. 1999; Vazquez \& Leitherer 2005). For our
application to the NGC 5253 cluster system, we adopted metallicities
of $Z=0.004$ and $0.008$ (where Z$_\odot = 0.019$). Adopting the
Calzetti et al. (2000) attenuation law, which is characterized by
total-to-selective absorption $R_V = 4.05$ and most appropriate for
use in starburst environments, we created a grid of models including
$0.0 \le E(B-V) \le 3.0$ mag for each age step.

For the gaseous contribution, our models employed standard H{\sc ii}
region values as input parameters to {\sc cloudy} (version 90.05;
Ferland et al. 1998), i.e., a hydrogen gas density of $10^2$
cm$^{-3}$, a filling factor of 0.01 and a covering factor of unity (no
leakage of ionizing photons, so that -- for instance -- all Lyman
continuum photons contribute to ionizing the gas). In addition, we
assumed that the metallicities of the stars and gas in a given cluster
are identical.

{\it Parameter determination.} We used a least-squares fit to
estimate, for each cluster, the best-fitting model (expressed in
magnitudes; see, e.g., Adamo et al. 2010a) and calculated the
associated $Q$ value (or $\chi^2$ probability function: models with $Q
\ga 0.1$ are good, while $Q > 0.001$ is still marginally acceptable;
Press et al. 1992), as well as the reduced $\chi^2_{\nu} =
\chi^2/\nu$, where $\nu = N - m - 1$ is the number of degrees of
freedom given $N$ data points and $m$ fit parameters. Here, $N$ refers
to the number of filters and $m = 2$, i.e., we obtain best fits of the
age and extinction ($m = 2$) and a scaled mass of each cluster for
fixed metallicity. The resulting 68 per cent confidence limits on the
determination of the cluster parameters were based on the prescription
of Lampton, Margon \& Bowyer (1976). Adamo et al. (2010a,b, 2011a,b)
used the spectral synthesis models of Marigo et al. (2008) to validate
their results.

Adamo et al. (2010a) discovered that several young clusters in the
luminous blue compact galaxy (BCG) Haro 11 have a flux excess in the
$I$, $H$ and $K_{\rm s}$ bands. Inclusion of those filters in the fit
resulted in significant residuals and a clear overestimation of the
cluster age. Given the starburst nature of NGC 5253, we expected that
at least some of the galaxy's YSCs might be affected by a similar red
excess.

Therefore, we produced three series of $\chi^2$ minimizations, i.e.,
one including all available data points from the NUV regime (F330W) to
the F222M filter, a second excluding F110W, F160W and F222M (i.e.,
from the NUV to the $I$ band), and a third which also excluded the $I$
band (leaving the range up to the F606W filter's effective
wavelength). Our initial analysis showed that the latter filter
selection did not produce any improvement to the fit (in the $\chi^2$
sense) and this filter selection leaves too few constraints on the
resulting cluster parameters because of a number of degeneracies
affecting the data, so that we discarded those results. 

We checked whether we could identify $I$-band excesses, but exclusion
of the $I$-band filter did not produce any improvement of the fit in
the $\chi^2$ sense (as expected if the fit including the $I$-band
photometry is correct). In this paper, we will thus only discuss the
NGC 5253 cluster fit results based on application of our approach to
(i) the full wavelength range and (ii) the range excluding the NIR
filters.

\subsection{GALEV models and AnalySED approach}
\label{galev.sec}

{\it Model-specific details.} Our second, complementary approach uses
the {\sc galev} SSP models (Kotulla et al. 2009; and references
therein, as well as subsequent unpublished updates). They cover ages
between $4 \times 10^6$ and $12.6 \times 10^{9}$ yr, with an age
resolution of $\Delta \log (t \mbox{ yr}^{-1}) = 0.05$, and are based
on the latest set of stellar isochrones of the Padova group for
metallicities of $0.0001 \le Z \le 0.03$, tabulated as 15 discrete
metallicities. Our model grid was completed by inclusion of extinction
effects (Calzetti et al. 2000), with $E(B-V)$ spanning the range from
0.0 to 2.0 mag, and adopting a resolution of 0.05 mag.

{\it Parameter determination.} In recent years, we developed a
sophisticated tool for star cluster analysis based on broad-band SEDs,
{\sc Analy\-SED}, which we tested extensively both internally (de
Grijs et al. 2003a,b; Anders et al. 2004) and externally (de Grijs et
al. 2005), using both theoretical and observed star cluster SEDs.

Note that in this paper we base our main analysis on the Yggdrasil
models and analysis approach discussed in Section \ref{redxs.sec}. We
apply the {\sc AnalySED} approach to the exact same filter set as used
in the IR-excess approach. This ensures that the results from both
approaches are directly comparable.

For a given cluster, the shape of the observed SED (again, expressed
in magnitudes), including the relevant observational uncertainties,
was compared with the shapes of model SEDs (as a function of age,
metallicity and foreground extinction) for the passband combination
considered. Each model SED (and its associated physical parameters)
was assigned a probability based on the $\chi^2$ value of this
comparison. The model data set with the highest probability (i.e., the
lowest $\chi^2$ value) was adopted as the most representative set of
cluster parameters. Models with decreasing probabilities were summed
up until a 68.26 per cent total probability (1$\sigma$ confidence
interval) was reached, to estimate the uncertainties in the
best-fitting model (cf. Anders et al. 2004). 

\section{Comparison of parameter determinations}
\label{robust.sec}

In the remainder of this paper, we will use our Yggdrasil models and
associated analysis, and the resulting fundamental cluster parameters,
as the basis for the ensuing discussion. In this section, we explain
in detail how we determined which clusters are affected by an IR
excess and pursue an approach in which we determine the basic cluster
parameters. In Sections \ref{sec4.1.sec} and \ref{sec4.2.sec}, we use
the Yggdrasil models and analysis as our basis; in Section
\ref{ageext.sec}, we compare these results with those from application
of the {\sc galev} models and the AnalySED approach.

\subsection{Determination of IR excesses}
\label{sec4.1.sec}

Based on application of our least-squares fitting routine to the full
wavelength coverage available (i.e., all 10 filters with central
wavelengths from 330 nm to 2.22 $\mu$m), we obtained best fits with
sufficiently small $\chi^2_{\nu}$ and $Q > 0.001$ for 149 of the 182
objects in our sample with photometric uncertainties $< 0.3$ mag in at
least 7 filters. In the remainder of this paper, we will base our
analysis on this reduced data set of 149 clusters and discard the 33
objects for which we could not achieve a satisfactory fit (we will
briefly return to these objects below). A number of clusters returned
best fits with $\chi^2_{\nu} \gg 1$ and $Q < 10^{-5}$. We inspected
the residuals, i.e., the differences between the observed SED and the
model fit, in each of the $N$ filters used for the fit (for all
clusters, independent of the goodness-of-fit parameters) and confirmed
that 30 objects were characterized by an IR excess with respect to the
best-fitting model that included all NUV and optical filters up to and
including F814W. The remaining 119 objects exhibited SEDs which could
be fitted with `standard' SSP models over the full wavelength range.

Fig. \ref{fig3.fig} shows comparisons of the resulting cluster
parameters. On the horizontal axes, we plot the ages, extinction
values and masses based on fits to the full F330W$\rightarrow$F222M
data set (10 filters), adopting a metallicity of $Z = 0.004$
henceforth. On the vertical axes, we plot the results for the
F330W$\rightarrow$F814W 7-filter set. The 30 clusters with appreciable
IR excess are shown as the objects contained in the (red)
circles. Overall, there is a reasonable one-to-one correlation between
the results from the two different data sets. As we already pointed
out in de Grijs et al. (2005), the mass determinations are clearly
better constrained than the age or extinction determinations. The
latter are very sensitive to small changes (including those due to
photometric problems) in the SEDs.

\subsection{Systematic effects for all versus IR-excess sources}
\label{sec4.2.sec}

Both the age and the mass comparisons for the sources {\it exhibiting
  an IR excess} show that exclusion of the NIR points results in
smaller values. For the mass determinations, this is expected at least
to some extent, because masses are determined based on a scaling of
the entire SED; excess intensity in a number of filters will hence
artificially increase the scaling upwards. In addition, redder SEDs
will result in older ages, and hence the resulting mass estimates will
be based on incorrect mass-to-light ratios, because the latter are
age-dependent. For the age determinations, the sources affected by an
IR excess are artificially given older ages (and older clusters are
assigned higher masses for the same luminosities), assuming fully
sampled cluster mass functions.

For the full sample of 149 objects, we found that the exclusion of the
NIR filters in our Yggdrasil-based analysis resulted in 112 objects
that exhibited an age difference of $\Delta \log t [{\rm yr}] \le
0.05$ dex. Eight objects were characterized by age differences of
$0.05 < \Delta \log t [{\rm yr}] \le 0.30$ dex, 18 objects resulted in
$0.30 < \Delta \log t [{\rm yr}] \le 0.50$ dex and for seven objects
$\Delta \log t [{\rm yr}] \ge 1.0$ dex. The equivalent numbers for the
differences in mass are 92 objects with $\Delta \log M_{\rm cl} [{\rm
    M}_\odot] \le 0.05$ dex, 27 exhibiting $0.05 < \Delta \log M_{\rm
  cl} [{\rm M}_\odot] \le 0.30$ dex, 16 for which $0.30 < \Delta \log
M_{\rm cl} [{\rm M}_\odot] \le 0.50$ dex and only three showing a
logarithmic age difference in excess of 1.0 dex.

We also checked whether our choice of metallicity may have affected
the results through the well-known age--metallicity degeneracy. We
refitted our entire data set, again using the two different filter
combinations, but now adopting $Z = 0.008$ for all clusters. The ages
resulting from the $Z = 0.008$ assumption were very similar to but
marginally younger than those based on $Z = 0.004$ (not shown). This
effect translated in a small to negligible offset that did not appear
to be a function of age. Thus, we conclude that our results are
negligibly affected by the age--metallicity degeneracy (but see
Section \ref{ageext.sec}), and that adoption of a single metallicity
for all clusters does not appreciably affect our analysis.

\begin{figure}
\psfig{figure=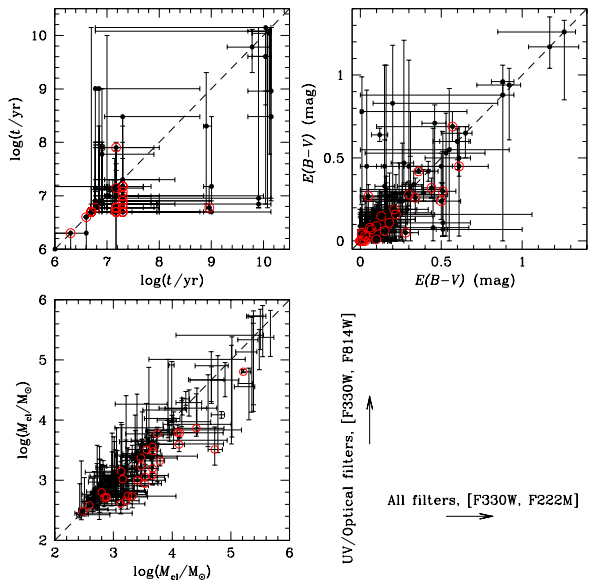,width=\columnwidth}
\caption{\label{fig3.fig}Comparison, using the models discussed in
  Section \ref{redxs.sec}, of the fundamental cluster parameters as a
  function of wavelength coverage. $X$ axis: all 10 filters; $y$ axis:
  all 7 NUV and optical filters, up to and including F814W. The
  objects identified with (red) circles are characterized by an IR
  excess (see text for discussion). Note that our standard SED
  analysis returns similarly young ages for many of the cluster
  candidates (cf. Fig. \ref{fig5.fig}, top left-hand
  panel). Therefore, the density of data points in the top left-hand
  panel appears lower than it is in reality.}
\end{figure}

\begin{figure}
\psfig{figure=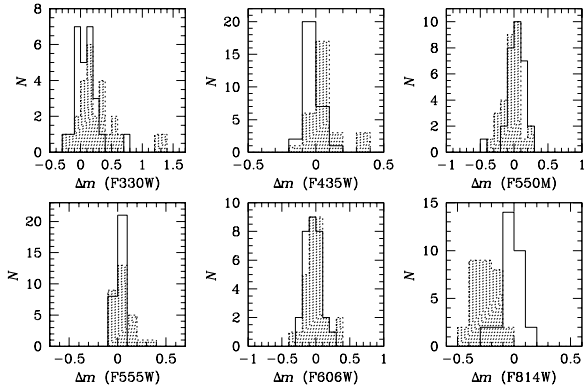,width=\columnwidth}
\vspace{-3cm}
\caption{\label{fig6.fig}Fit residuals as a function of filter for the
  clusters affected by a significant IR excess. The shaded histograms
  are the residuals based on fits to the full available wavelength
  range; the solid histograms show the results after exclusion of the
  filters affected by the IR excess.}
\end{figure}

In Fig. \ref{fig6.fig} we compare the residuals of the fits obtained
from single-mindedly fitting the entire available wavelength range
(shaded histograms) and those resulting after exclusion of the NIR
passbands (open histograms drawn using solid lines), again for the
clusters affected by an IR excess only. It is clear that the residuals
in the latter case are more symmetrically distributed around
zero. Note that the most significant effect is seen in the residuals
of the F814W filter. These residuals improve upon exclusion of the NIR
data because this allows us to obtain a `proper' match (in the
$\chi^2$ sense) of all data points in the optical regime. Because of
the flux excess in the NIR filters, in the fits including all 10
filters the observed F814W flux was consistently below the best model
prediction.

\begin{figure}
\psfig{figure=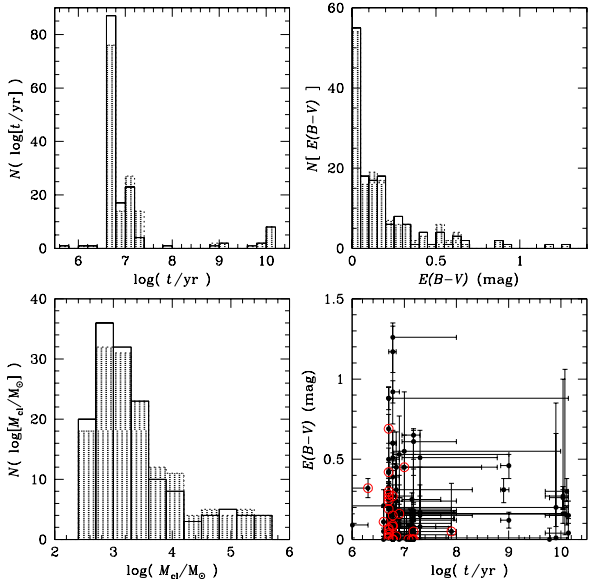,width=\columnwidth}
\caption{\label{fig5.fig}Distributions of the fundamental NGC 5253
  cluster parameters. The dotted histograms with dashed shading
  represent the parameters based on fits to the full wavelength range
  from the F330W to the F222M filter; the histograms drawn using solid
  lines are the resulting distributions after removal of the NIR
  filters from the SED analysis. The bottom right-hand panel shows the
  distribution of our sample clusters in the age--extinction plane
  (see Section \ref{sizes.sec}). The objects identified with red
  circles are characterized by an IR excess.}
\end{figure}

Fig. \ref{fig5.fig} is a collection of the distributions of the basic
cluster parameters for the full data set of 149 clusters. We show both
the distributions initially obtained by fitting the entire available
wavelength range (dotted histograms) and those resulting from NUV to
optical filters only (solid lines). As in Fig. \ref{fig3.fig}, we see
that this filter selection leads to younger ages and lower masses,
while the overall distribution of extinction values remains
statistically unchanged. Note that the large majority of our sample
clusters, particularly those exhibiting an IR excess (see also
Fig. \ref{fig3.fig}), have masses between a few $\times 10^2$ and
$10^5$ M$_\odot$ (for confirmation of this mass range, see e.g.,
Harris et al. 2004; Cresci et al. 2005). These are most likely
significantly affected by stochastic sampling of their stellar mass
functions. Table \ref{Yggdrasil.tab} includes our final set of derived
cluster parameters based on the Yggdrasil SSP models and associated
analysis (assuming fully sampled mass functions).

\begin{table*}
\caption[ ]{\label{Yggdrasil.tab}NGC 5253 cluster coordinates and
  derived parameters assuming fully sampled Yggdrasil SSP models.$^a$}
\begin{center}
{\scriptsize
\begin{tabular}{ccccccccccc}
\hline
\hline
ID & R.A. (J2000) & Dec. (J2000) & $\sigma_{\rm G, F550M}$ & $N_{\rm filters}$$^b$ & $\log( t )$ & \multicolumn{2}{c}{Uncertainties} & $\log( M_{\rm cl} )$ & \multicolumn{2}{c}{Uncertainties} \\
\cline{7-8}\cline{10-11}
   & (deg)        & (deg)        & (pixels)  & & [yr] & (positive) & (negative) & [M$_\odot$] & (positive) & (negative) \\
\hline
0  & 204.9766060 & $-$31.6426963 & 1.154 & $\cdots$ & $\cdots$ & $\cdots$ & $\cdots$ & $\cdots$ & $\cdots$ & $\cdots$\\
1  & 204.9775826 & $-$31.6427056 & 1.165 & 10 & 7.18 & 0.67 & 0.40 & 3.13 & 0.26 & 0.61 \\
2  & 204.9810049 & $-$31.6366573 & 1.194 & 10 & 7.15 & 0.03 & 0.37 & 3.04 & 0.13 & 0.46 \\
3  & 204.9810436 & $-$31.6368153 & 1.158 & 10 & 9.79 & 0.33 & 0.48 & 4.67 & 0.21 & 0.39 \\
4  & 204.9792423 & $-$31.6411761 & 1.154 & 10 & 6.78 & 0.03 & 0.03 & 2.71 & 0.10 & 0.03 \\
$\cdots$ & $\cdots$ & $\cdots$ & $\cdots$ & $\cdots$ & $\cdots$ & $\cdots$ & $\cdots$ & $\cdots$ & $\cdots$ & $\cdots$ \\
\hline
\end{tabular}
}
\end{center}
\flushleft $^a$ The full data table is available associated with the
online version of this article. This stub is meant to provide guidance
regarding form and content of the full table. $^b$ Number of filters
used to obtain the derived parameters.
\end{table*}

\begin{figure}
\psfig{figure=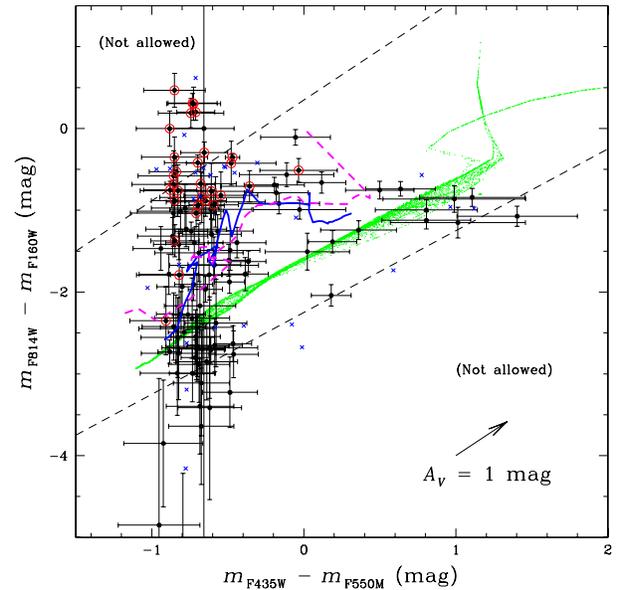,width=\columnwidth}
\caption{\label{fig8.fig}Colour--colour diagram in the ST mag
  system. Blue heavy solid versus magenta heavy dashed line: {\sc
    galev} versus Yggdrasil SSP models for $Z = 0.004$, spanning an
  age range from 3 Myr (blue extremeties) to $\sim 12$ Gyr (red
  extremeties of the models). Green points: Theoretical stellar loci
  (Marigo et al. 2008); small blue crosses: Objects for which we could
  not obtain a satisfactory fit using fully sampled SSP models. The
  objects identified with red circles are characterized by an IR
  excess. The black dashed line delineates the region in
  colour--colour space where star clusters and individual stars may be
  found (taking into account realistic photometric uncertainties).}
\end{figure}

We also explored whether our results are affected significantly by the
age--extinction degeneracy, which often causes artificial offsets if
one has access to only broad-band SEDs covering smaller wavelength
ranges. The optical--NIR colour--colour diagram of Fig. \ref{fig8.fig}
clearly shows that the clusters in which we detected an IR excess are
all located redward of the model SSP (blue solid line). This,
therefore, serves as an independent check of the reality of our
IR-excess determinations (see also fig. 2 of Cresci et al. 2005).

The large majority of IR-excess sources are located towards blue
optical colours. If they were affected by significant extinction, we
would have expected them to be displaced well along the direction of
the reddening vector. However, since we have access to up to 10 data
points spanning the entire NUV/optical/NIR range, this helps our
efforts to break the age--extinction degeneracy quite significantly:
for instance, the evolution in age and the extinction vector are not
oriented parallel to each other in the colour--colour diagram of
Fig. \ref{fig8.fig}, at least for young ages (they would be in
optical-only colour--colour diagrams). We thus conclude that the young
IR-excess sources are, by and large, negligibly affected by the
age--extinction degeneracy and, more importantly, that the IR excess
in these sources is most likely not owing to the effects of
extinction. For reference, we also included the 33 objects for which
we could not obtain satisfactory fits using fully sampled SSP models
in Fig. \ref{fig8.fig} (small blue crosses). Their distribution in
colour--colour space follows that of the remaining 149 objects which
are the main focus of this paper; a fraction also appears to exhibit
an IR excess with respect to the standard SSP models.

\begin{figure}
\psfig{figure=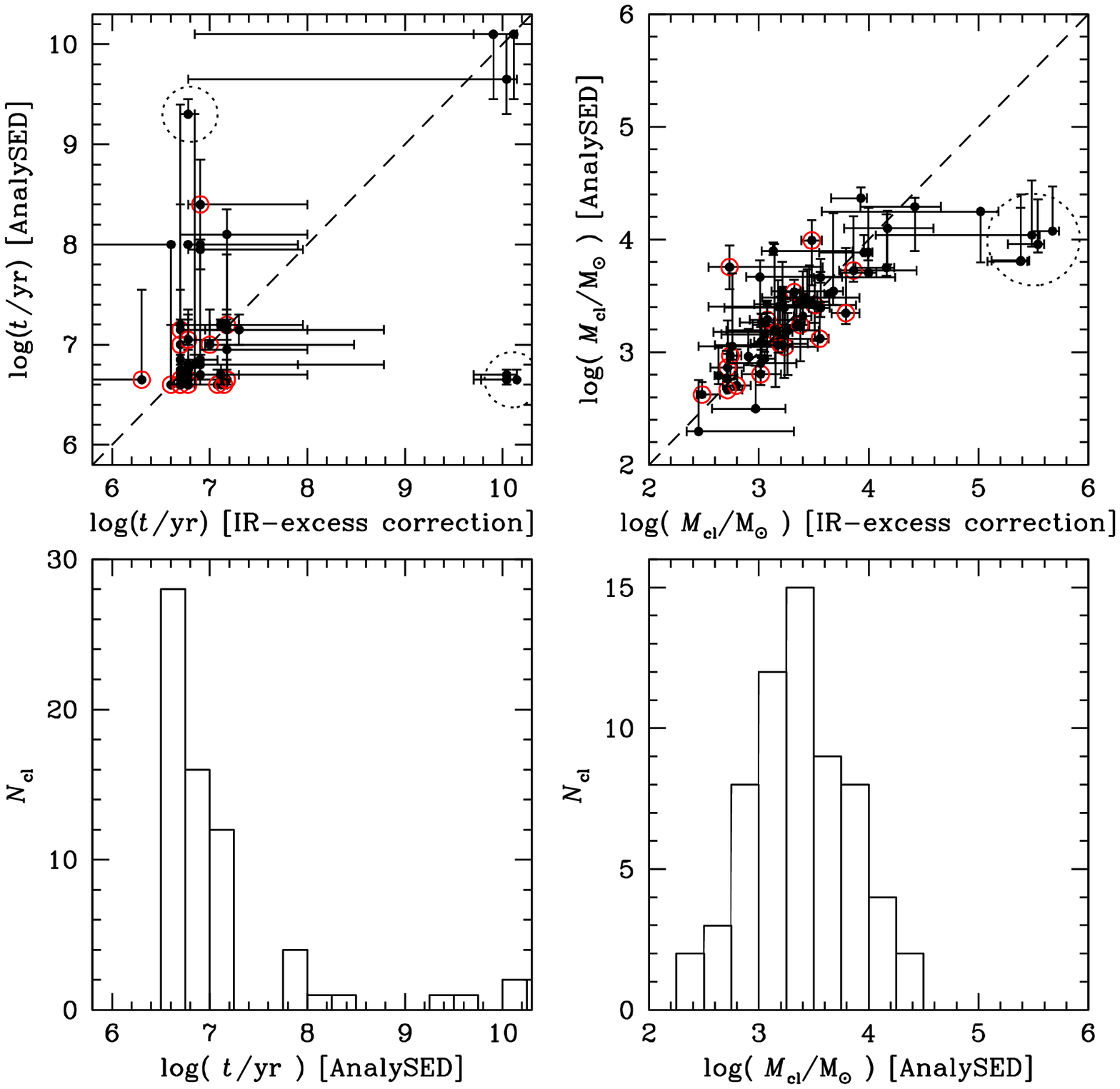,width=\columnwidth}
\caption{\label{fig7.fig}{\it (top)} Comparison of the individual
  cluster ages and masses based on application of our two different
  SSP-matching approaches. The objects identified with (red) circles
  are characterized by an IR excess. The dotted circles indicate the
  handful of sources which are most likely significantly affected by
  the age--extinction degeneracy (see Section \ref{ageext.sec}). {\it
    (bottom)} Age and mass distributions based on application of the
  {\sc AnalySED} approach. }
\end{figure}

\subsection{One-to-one comparisons: dependence on the SSP model suite adopted}
\label{ageext.sec}

Finally, we compare the resulting masses and ages from our Yggdrasil
models and analysis with those obtained based on the {\sc galev}+{\sc
  Analy\-SED} approach: see Fig. \ref{fig7.fig} and Table
\ref{galev.tab}. The top row of Fig. \ref{fig7.fig} shows the
one-to-one comparisons of the clusters' ages and masses (the resulting
extinction values cannot be determined as reliably). While the overall
trends show reasonable agreement, particularly for the masses (cf. de
Grijs et al. 2005) as well as for most of the IR-excess sources, there
are a few discrepancies (indicated by the dotted circles) that require
further investigation.

\begin{table*}
\caption[ ]{\label{galev.tab}NGC 5253 cluster parameters derived
  assuming fully sampled {\sc galev} SSP models.$^a$}
\begin{center}
\begin{tabular}{ccccccc}
\hline
\hline
ID & $\log( t )$ & \multicolumn{2}{c}{Uncertainties} & $\log( M_{\rm cl} )$ & \multicolumn{2}{c}{Uncertainties} \\
\cline{3-4}\cline{6-7}
   & [yr] & (positive) & (negative) & [M$_\odot$] & (positive) & (negative) \\
\hline
0  & $\cdots$ & $\cdots$ & $\cdots$ & $\cdots$ & $\cdots$ & $\cdots$ \\
1  & 6.65 & 0.05 & 0.05 & 2.62 & 0.11 & 0.03 \\
2  & 6.65 & 0.55 & 0.05 & 3.24 & 0.18 & 0.32 \\
3  & 6.60 & 0.15 & 0.05 & 2.70 & 0.32 & 0.03 \\
4  & 7.00 & 0.35 & 0.05 & 3.73 & 0.48 & 0.10 \\
$\cdots$ & $\cdots$ & $\cdots$ & $\cdots$ & $\cdots$ & $\cdots$ & $\cdots$ \\
\hline
\end{tabular}
\end{center}
\flushleft $^a$ The full data table is available associated with the
online version of this article. This stub is meant to provide guidance
regarding form and content of the full table.
\end{table*}

\begin{figure*}
\begin{tabular}{cc}
\psfig{figure=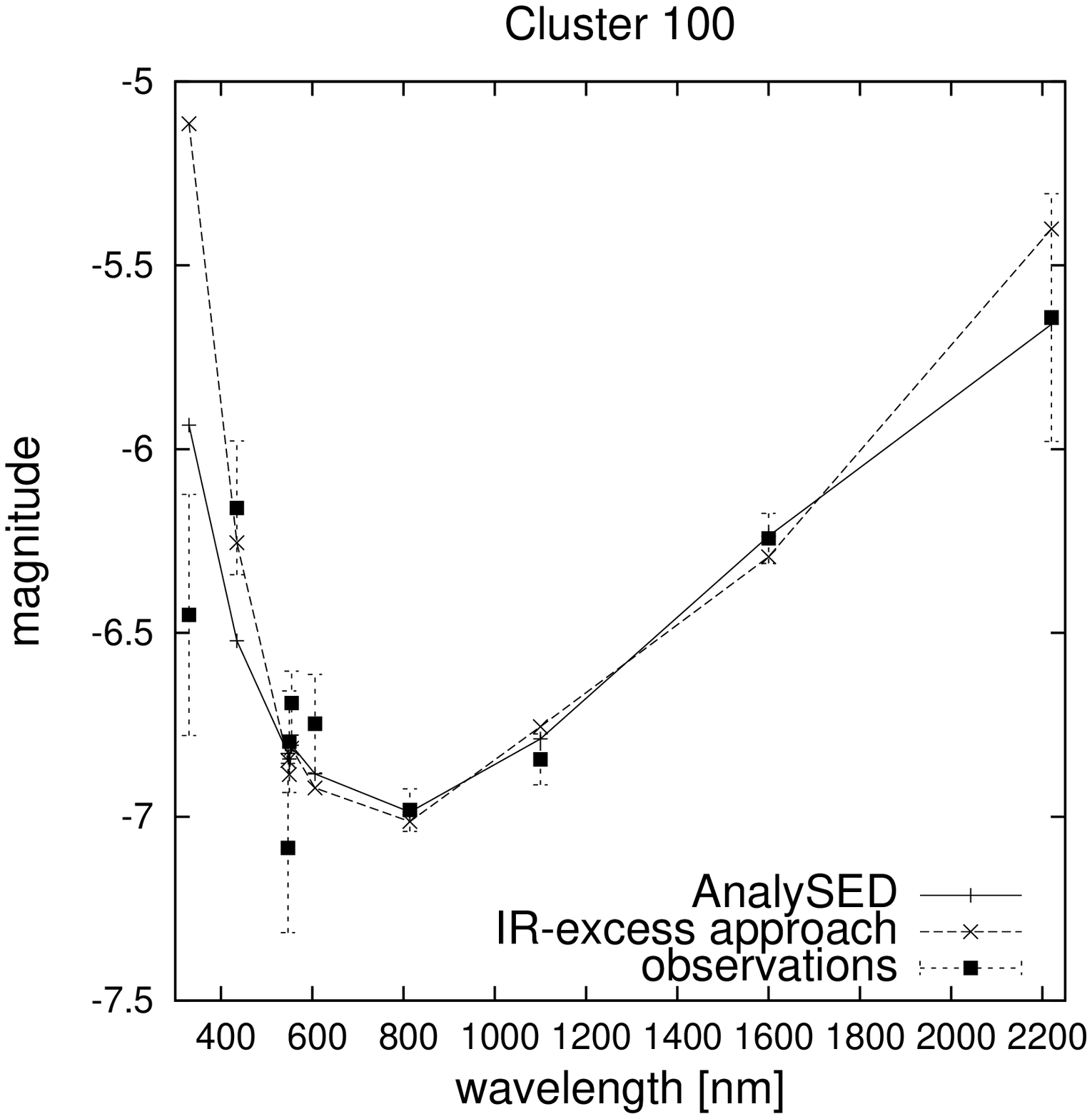,width=0.65\columnwidth,angle=-90} &
\psfig{figure=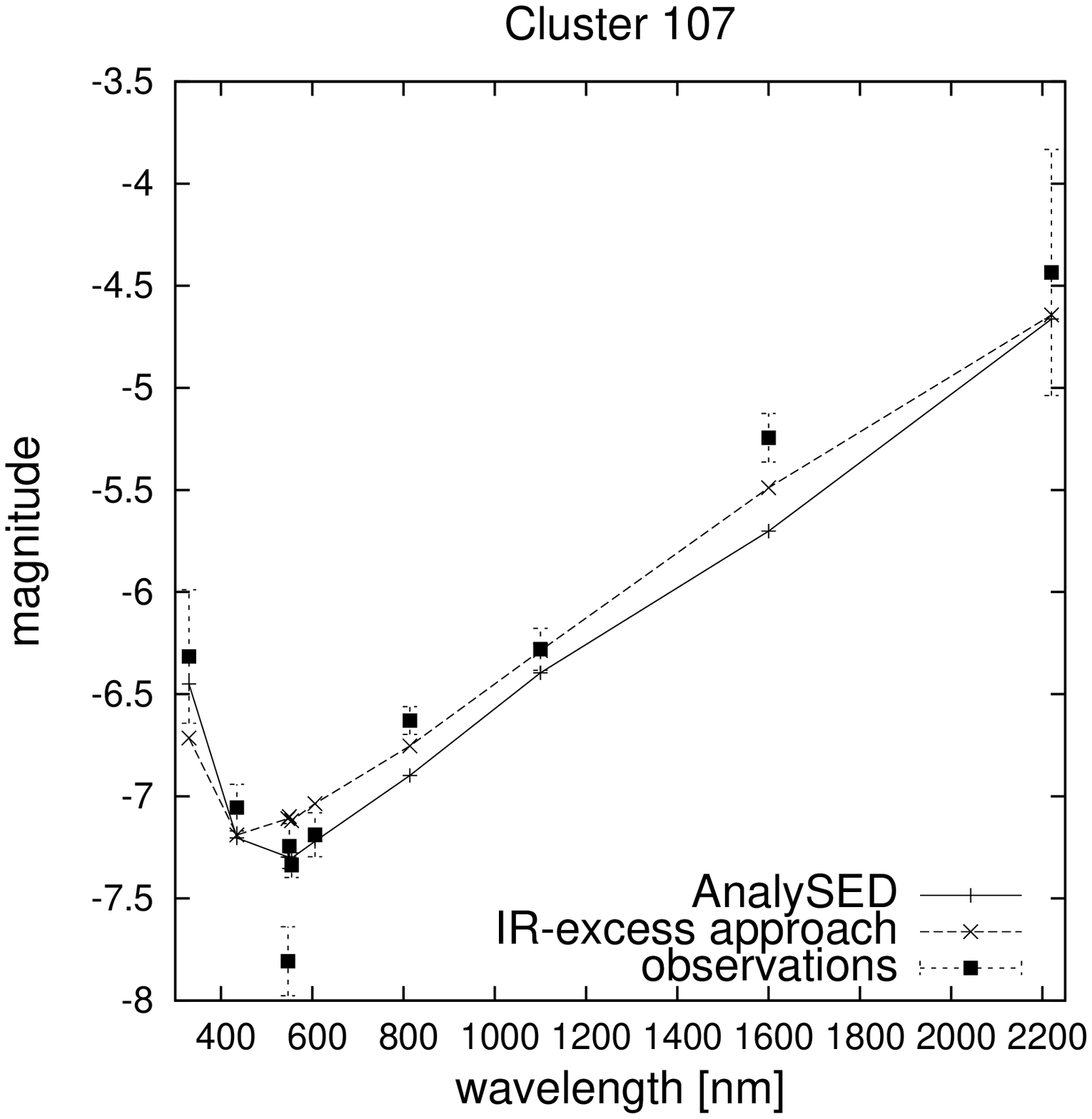,width=0.65\columnwidth,angle=-90} \\
\psfig{figure=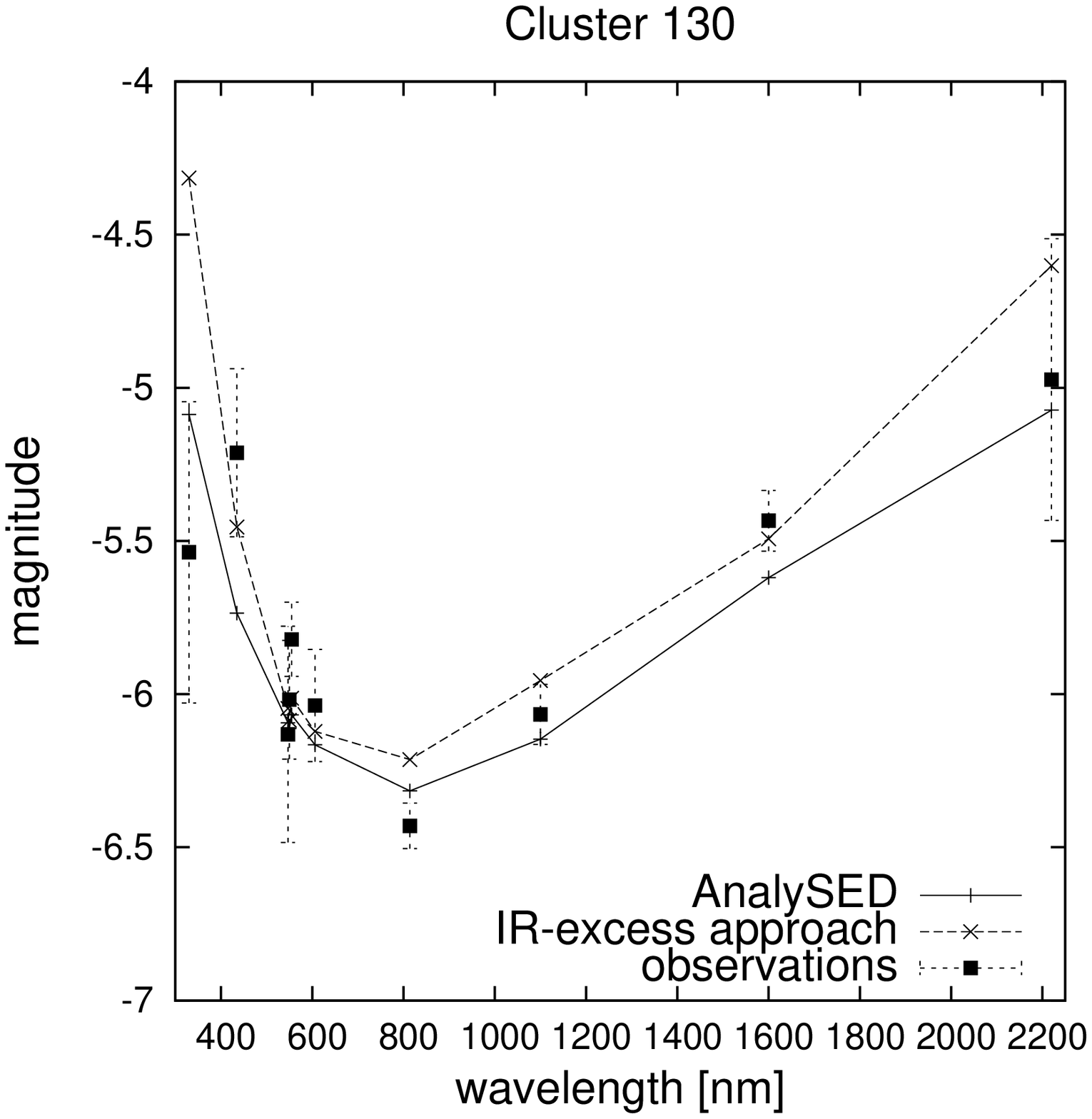,width=0.65\columnwidth,angle=-90} &
\psfig{figure=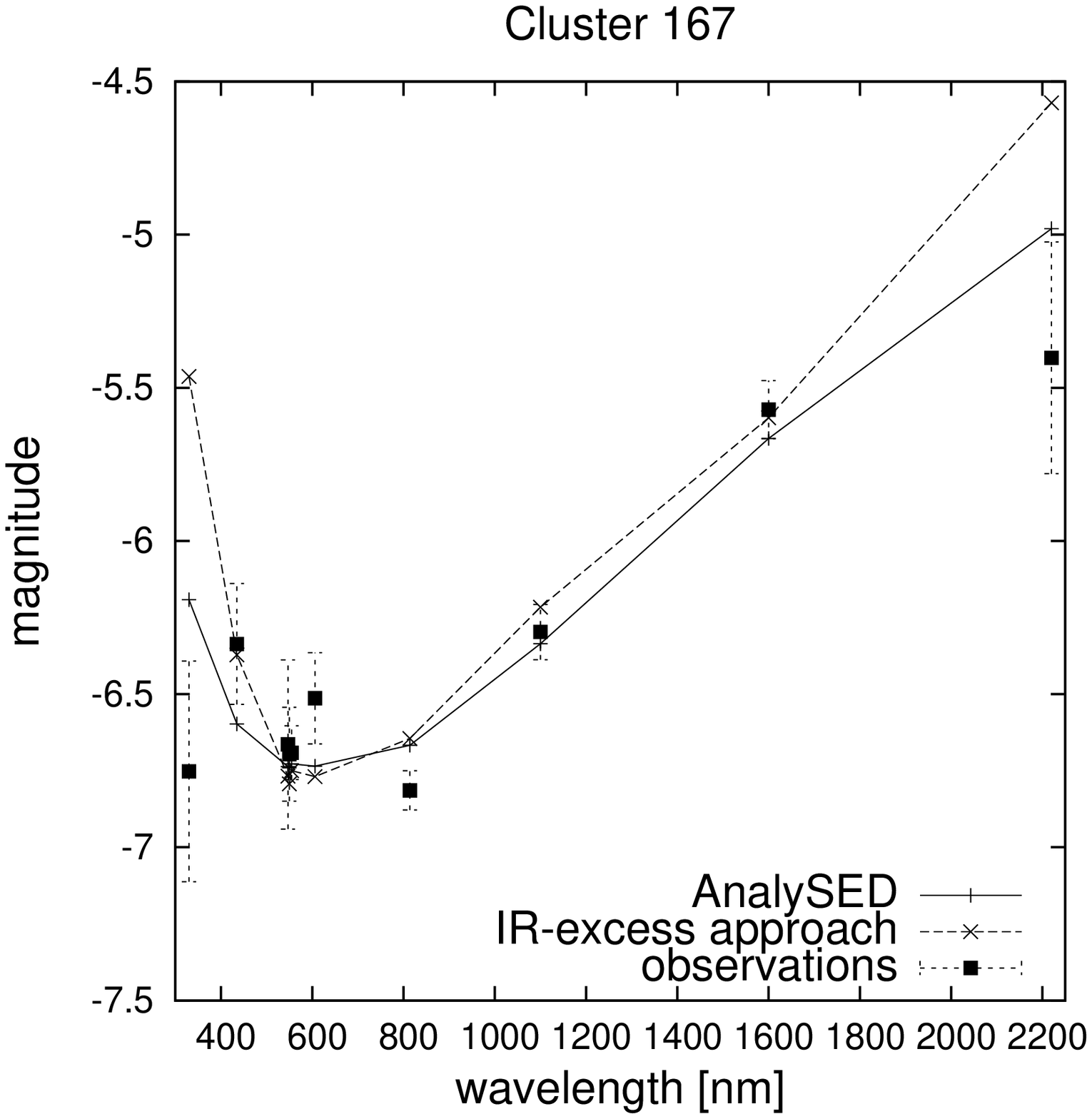,width=0.65\columnwidth,angle=-90} \\
\end{tabular}
\caption{\label{discrepancies.fig}SEDs (absolute magnitudes in the ST
  mag system) of clusters 100, 107, 130 and 167 (contained in the
  dotted circles in Fig. \ref{fig7.fig}), showing the effects of the
  age--extinction degeneracy (see text). The model SEDs are all based
  on {\sc galev} models, adopting the best-fitting ages and extinction
  values derived from the {\sc AnalySED} and Yggdrasil-based
  (`IR-excess') approaches.}
\end{figure*}

Fig. \ref{discrepancies.fig} displays the SEDs of the four clusters
for which our Yggdrasil-based analysis and the {\sc AnalySED} approach
obtained the most different age estimates [$\Delta t$ (yr) $> 2
  \sigma$]. The latter returned an age for cluster 107 (indicated by
the small dotted circle in the top left-hand panel of
Fig. \ref{fig7.fig}) of $2.0 \pm 0.8$ Gyr and an extinction value of
$E(B-V) = 0.0$ mag, while our Yggdrasil-based analysis resulted in
estimated age and extinction values of $6.0 \pm 0.7$ Myr and 0.60 mag,
respectively, for similar-quality fits. Clusters 100, 130 and 167
(contained in the large dotted circle in the same panel) are also
clearly discrepant because of the age--extinction degeneracy: the
(age, extinction) pairs for each of these clusters are, respectively,
(4.5 Myr, 1.05 mag) versus (11.2 Gyr, 0.25 mag), (5.0 Myr, 1.10 mag)
versus (11.2 Gyr, 0.25 mag), and (4.5 Myr, 0.90 mag) versus (12.6 Gyr,
0.05 mag), where the first and second (age, extinction) pairs refer to
the results from the {\sc AnalySED} and Yggdrasil-based analyses,
respectively. Additionally, differences between both sets of models
and the details of the fitting routines may contribute to a lesser
extent. The significantly different age estimates for these four
clusters are associated with similarly varying mass-to-light ratios,
which in turn also lead to significant differences in the resulting
mass estimates. This effect is seen in the top right-hand panel of
Fig. \ref{fig7.fig}, where the objects of relevance are indicated by
the large dotted circle.

Despite these differences for a small number of clusters, the main
features in the overall age and mass distributions (bottom row) are
fairly well reproduced by both independent approaches. Assuming a
simple underlying skewed Gaussian distribution, the peak in the age
distribution resulting from our {\sc AnalySED} analysis (including all
objects) occurs at $\langle \log(t \mbox{ yr}^{-1}) \rangle = 7.11$,
with $\sigma_{\langle \log t \rangle} = 0.80$. For comparison, the
Yggdrasil-based analysis returned $\langle \log(t \mbox{ yr}^{-1})
\rangle = 7.10$, with $\sigma_{\langle \log t \rangle} = 0.94$. For
the mass distributions, we find $\langle \log(M_{\rm cl}/{\rm
  M}_\odot) \rangle = 3.38$ versus 3.37, and $\sigma_{\langle \log
  M_{\rm cl} \rangle} = 0.76$ versus 0.77, respectively. This
excellent correspondence is much better than the feature-reproduction
robustness reported in de Grijs et al. (2005), where we quoted
reproducibilities of $\Delta \langle \log(t \mbox{ yr}^{-1})\rangle
\le 0.35$ and $\Delta \langle \log( M_{\rm cl} / {\rm M}_\odot)\rangle
\le 0.14$.

If we compare the age and mass determinations for the individual
clusters, a similar result emerges. The mean differences between the
Yggdrasil-based analysis and the {\sc AnalySED} results is $\langle
\Delta \log(t \mbox{ yr}^{-1})\rangle = 0.041$ and $\langle \Delta
\log( M_{\rm cl} / {\rm M}_\odot)\rangle = 0.100$, respectively, with
spreads (Gaussian $\sigma$'s) of 0.900 and 0.517 dex, respectively.

\subsection{To which extent does H$\alpha$ photometry make a difference?}
\label{halpha.sec}

So far, we have focussed on the power of (medium- and) broad-band SED
analysis based on extensive wavelength coverage. This is the type of
observational data one might expect to obtain most easily for any
extragalactic star cluster system, and hence exploration of the
limitations of such a data set is relevant. For our analysis of the
specific star cluster sample of NGC 5253, we also obtained narrow-band
F658N observations from the {\sl HST} Data Archive, covering the
H$\alpha$ emission line.\footnote{Note that NGC 5253 was also observed
  through a narrow-band filter covering the H$\beta$ line. Calzetti et
  al. (1997) already extensively explored the galaxy's morphology and
  dust properties based on these observations, so that we will not
  repeat that analysis here.} For most of the NGC 5253 clusters,
numerous authors, including Calzetti et al. (1997) and Cresci et
al. (2005), provide support for the young ages resulting from their
fits by referring to the H$\alpha$ fluxes found associated with these
objects. H$\alpha$ fluxes have also been found useful for analyses of
numerous other YSC systems (for recent references, see e.g., Chandar
et al. 2011; Whitmore et al. 2011; Fouesneau et al. 2012; and
references therein), although the present analysis represents the
first full integration of H$\alpha$ photometry in a simultaneous
multi-passband ($N > 4$ filters) approach.

In this section, we add the F658N/H$\alpha$ photometry to our SEDs for
confirmation and reanalysis of the properties of the NGC 5253 clusters
using the Yggdrasil models. We obtained two sets of F658N photometry
for our star cluster sample. First, we adopted the same apertures as
for the nearest continuum filter, F550M. However, H$\alpha$ emission
may be more extended than that of the associated continuum sources. As
such, we determined the Gaussian $\sigma_{\rm G}$'s on the F658N image
itself and used these size estimates to determine a second set of
F658N magnitudes (using the source, inner and outer sky radii
expressed in units of $\sigma_{\rm G}$ as defined in Section
\ref{photom.sec}). A comparison of Gaussian sizes (not shown)
indicates that the H$\alpha$ emission coincident with our sample
clusters exhibits a much broader range than the associated continuum
emission in the F550M filter.

\begin{figure}
\psfig{figure=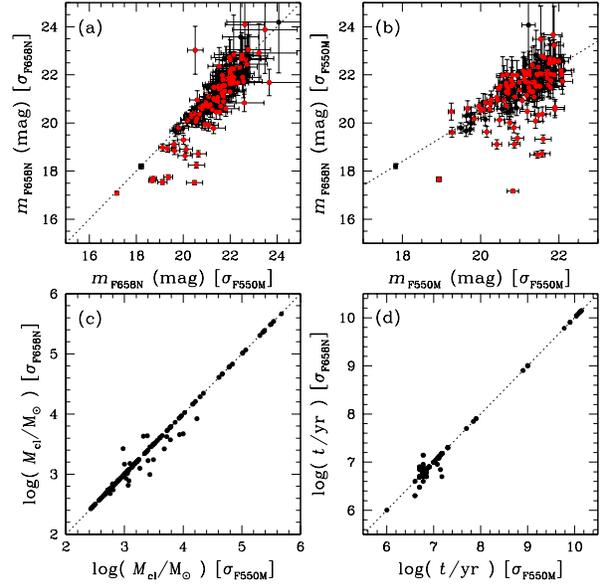,width=\columnwidth}
\caption{\label{halpha.fig}(a) F658N photometry as a function of
  aperture size based on the intrinsic Gaussian $\sigma_{\rm G}$'s in
  the F550M and F658N images using the individualized apertures
  defined in Section \ref{photom.sec}. Red data points: objects
  younger than 10 Myr; dotted line: locus of equality. (b) Cluster
  photometry in the F550M and F658N filters, using aperture sizes
  based on $\sigma_{\rm G, F550M}$. Dotted line: locus of equality
  corrected for the difference in filter transmission windows. (c) and
  (d) Mass and age comparisons, respectively, obtained using the
  photometry based on the differently sized apertures from panel
  (a). We have left out the relevant error bars for presentational
  clarity. All panels include the data or derived properties for all
  149 clusters in our final sample.}
\end{figure}

Fig. \ref{halpha.fig}a shows the resulting F658N photometry as a
function of aperture size used. Although a reasonable one-to-one
correlation between the F658N magnitudes resulting from using aperture
sizes based on the F550M and F658N $\sigma_{\rm G}$'s is apparent,
there is a clear offset to brighter magnitudes if the F658N-based size
criterion is used, particularly for objects returned as younger than
10 Myr, based on the Yggdrasil-based broad-band analysis (indicated in
red). This underscores the need to account for differences in source
sizes as a function of passband.

Panel (b) shows the distribution of our sample clusters in the F550M
versus F658N magnitude plane (the latter includes both continuum and
line flux, if present). Again, the red data points are the objects
younger than 10 Myr and the dotted line is the locus of equality,
corrected for the difference in filter transmission windows. This
panel shows that all sources that are more than twice their
photometric $1\sigma$ uncertainties from the dotted line are young ($<
10$ Myr). Note that we used the same apertures here for both axes, so
that we can properly compare the presence of any H$\alpha$ excess.

Figs \ref{halpha.fig}c and d compare the masses and ages,
respectively, obtained using the photometry based on the differently
sized apertures shown in panel (a). We have left out the error bars
for presentational clarity (see, e.g., Fig. \ref{fig3.fig} for
guidance). These panels show that the impact of the different aperture
sizes is very small, even at ages where significant H$\alpha$ emission
may be expected.

\begin{figure}
\psfig{figure=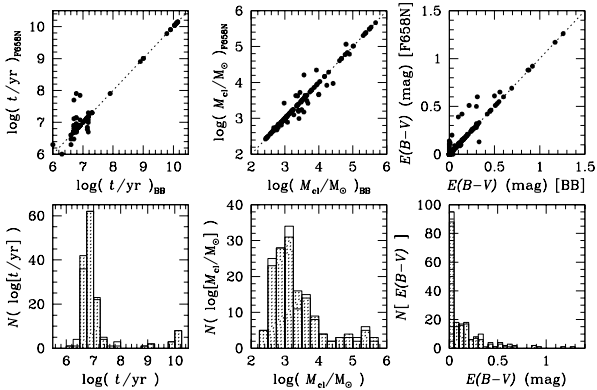,width=\columnwidth}
\vspace{-3cm}
\caption{\label{halpha2.fig} (top row) One-to-one comparisons for the
  full sample of 149 clusters of the ages, masses and extinction
  values based on using our medium- and broad-band photometry only
  (`BB'), and that based on also including the F658N data. (bottom
  row) Dotted histograms: distributions based on the medium- and
  broad-band data only; solid histograms: based on additional
  inclusion of the F658N fluxes.}
\end{figure}

In Fig. \ref{halpha2.fig} (top row) we compare the ages, masses and
extinction values based on using our (medium- and) broad-band
photometry only (`BB'), and that based on also including the F658N
data. The bottom row shows the equivalent histograms of these
parameters. The dotted histograms are the distributions based on the
(medium- and) broad-band data only; the solid histograms are based on
additional inclusion of the F658N fluxes. The differences, in any of
the panels, are very small and statistically negligible. This is
exemplified by a quantitative comparison of the histograms'
characteristics. The peak and Gaussian width of the age distribution
resulting from our analysis including the F658N photometry are
$\langle \log (t \mbox{ yr}^{-1} ) \rangle = 7.13$ and
$\sigma_{\langle \log t \rangle} = 0.94$ dex, respectively, compared
with 7.10 and 0.94 dex for the analysis based on the (medium- and)
broad-band images only. Similarly, the mean and Gaussian width of the
equivalent mass distributions are $\langle \log(M_{\rm cl}/{\rm
  M}_\odot) \rangle = 3.36$ and $\sigma_{\langle \log M_{\rm cl}
  \rangle} = 0.77$ dex, respectively, versus 3.37 and 0.76 dex. We
thus conclude that the H$\alpha$ photometry fully supports our age and
mass determinations of the NGC 5253 cluster sample based on medium-
and broad-band photometry alone.

\section{Initial assessment of the effects of stochasticity}
\label{stoch.sec}

Stochastic sampling of the stellar mass function, particularly of the
massive stars in relatively low-mass ($\la$ a few $\times 10^4$
M$_\odot$) -- and therefore relatively poorly populated -- clusters,
may significantly affect the determination of fundamental cluster
parameters based on broad-band SEDs (e.g., Cervi\~no, Luridiana \&
Castander 2000; Cervi\~no et al. 2002; Cervi\~no \& Luridiana 2004,
2006; Barker et al. 2008; Ma\'{\i}z Apell\'aniz 2009; Popescu \&
Hanson 2010; Fouesneau \& Lan\c{c}on 2010; Fouesneau et al. 2012;
Popescu, Hanson \& Elmegreen 2012). In general, stochastic sampling of
cluster stellar mass functions leads to broad, asymmetric colour
distributions compared with the equivalent distributions associated
with fully sampled SSP models and, as a consequence, important local
biases are introduced in cluster age determinations for masses $M_{\rm
  cl} \la$ a few $\times 10^4$ M$_\odot$. The resulting cluster {\it
  masses} are usually less affected by stochastic sampling than the
corresponding age estimates; they tend to scatter around the
equivalent mass determinations derived using the traditional approach.

Inspection of the derived cluster masses in Figs \ref{fig5.fig} and
\ref{fig7.fig} clearly indicates that the majority of our sample
clusters may be affected by stochastic sampling of their mass
functions (cf. the mass ranges quoted in Harris et al. 2004; Cresci et
al. 2005). This observation is further compounded by the results of
our age determinations (based on the assumption of fully sampled
SSPs!), which imply that a large fraction of the NGC 5253 clusters
have ages near $10^7$ yr (as confirmed and tightly constrained by the
H$\alpha$ analysis in Section \ref{halpha.sec}).

This corresponds to stellar population ages when luminous red stars
are present. Red-supergiant stars (RSGs; with stellar masses up to
$\sim 40$ M$_\odot$), in particular, start to appear around this time
in realistic stellar populations, while at the same time the massive,
hot, ionizing stars that dominate at young ages may cause significant
variations in the nebular emission from stochastically sampled
clusters (e.g., Cervi\~no et al. 2000; Fouesneau et al. 2012). These
stars are significantly brighter than their main-sequence
counterparts, but not proportionally more massive. The absence or
presence of a small number of such stars can significantly affect the
observed integrated colours of a cluster. For instance, stochastic
effects are expected to affect the colour and magnitude of a 10
Myr-old, $10^5$ M$_\odot$ cluster by $\Delta V = 0.08$ and $\Delta
(V-I) = 0.07$ mag (Dolphin \& Kennicutt 2002). In addition, these
sources are relatively short-lived, and so the luminosity of the
clusters might be expected to vary significantly on short timescales
(e.g., Kiss et al. 2006; Szczygie{\l} et al. 2010; and references
therein). This is of particular interest and relevance in the present
context because our observational data were taken over a period
spanning a total of $\sim 11$ years between the first and final images
included in this study (proposal IDs 5479 and 10609, respectively).

We can obtain a better (initial) handle on the significance of the
stochastic sampling effects associated with our cluster sample by
examining Fig. \ref{fig8.fig} in detail. First, we note that most of
our clusters are young and relatively unaffected by extinction. For
$A_V = 0$ mag, it appears that most of the clusters can be explained
by a young age ($\la 10^8$ yr). However, in this case neither clusters
redder than $(m_{\rm F435W} - m_{\rm F550M}) = 0.4$ mag, nor those
redder than the $(m_{\rm F814W} - m_{\rm F160W})$ values defined by
the SSP models can be explained. This implies that at least some
amount of extinction is required to match the distribution of the
clusters in colour--colour space. In addition, clusters that are
redder in $(m_{\rm F814W} - m_{\rm F160W})$ than the envelope of the
SSP models with variable extinction are most likely explained by the
presence of individual stars. This includes the majority of objects we
identified as having an IR excess. (Note that we cannot exclude the
possibility that, in a small number of cases, an additional photon
source that is unaccounted for may also give rise to their violation
of the observational boundaries for genuine stars and SSPs.)

As such, we conclude that the IR excesses observed in 30 of the NGC
5253 clusters are consistent with the effects of stochastic
sampling. Similarly, the additional clusters found in the same part of
the colour--colour diagram are likely also affected by stochastic
sampling if we adopt the same reasoning. This includes 30--40 per cent
of the total sample discussed in this paper.

A first investigation focussing on the accuracy of derived cluster
ages obtained along similar lines as done in this paper was published
by Silva-Villa \& Larsen (2011). We are in the process of implementing
direct stochastic modelling using the {\sc galev} SSPs as our basis. A
preliminary comparison of our models with the results of Silva-Villa
\& Larsen (2011) indicates good agreement (E. Silva-Villa,
priv. commun.). We note, however, that we are introducing important
improvements with respect to their work. In particular, our
observations of the NGC 5253 cluster system cover a much larger
wavelength range, while we are also investigating the effects of
extinction as a free parameter in our fits. A much more in-depth
investigation of the effects of stochastic sampling on the ages and
masses determined based on broad-band photometry will be published in
Paper II. Once we have fully validated our stochastic modelling
approach, we will undertake a detailed study of the effects of
stochastic sampling on the derivation of the NGC 5253 fundamental
cluster parameters (Paper III). This will have important consequences
for our understanding of the system, given that many clusters have
masses in the regime that is dominated by stochastic sampling
effects. The current cluster parameters will then provide a very
useful comparison data set.

\section{Physical state of the NGC 5253 cluster population}
\label{overall.sec}

\subsection{The cluster age distribution}
\label{agedist.sec}

\begin{figure}
\psfig{figure=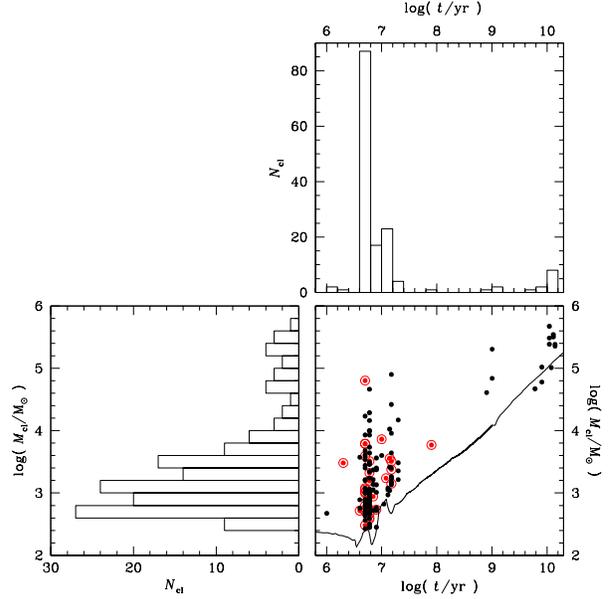,width=\columnwidth}
\caption{\label{agemass.fig}Age--mass distribution of the NGC 5253
  clusters. The objects identified with (red) circles are
  characterized by an IR excess. The solid line in the age--mass plane
  represents our average, galaxy-wide detection limit at $M_V \simeq
  -5.4$ mag ($V \simeq 22.1$ mag), based on the SSP models
  adopted. For reasons of presentational clarity, we do not include
  error bars on the individual data points.}
\end{figure}

Fig. \ref{agemass.fig} shows the distribution of our final,
well-fitted NGC 5253 star cluster sample (of 149 objects) in the
diagnostic age--mass plane, as well as the age and mass distributions
resulting from our IR-excess analysis. The cluster population is
dominated by a significant number of relatively low-mass ($M_{\rm cl}
\ll 10^5$ M$_\odot$) objects. Assuming fully sampled stellar mass
functions, these objects have ages from a few $\times 10^6$ to a few
$\times 10^7$ yr. This age range is in excellent agreement with the
starburst age of the host galaxy, $\le 10$ Myr, and is confirmed by
our age redeterminations based on the clusters' H$\alpha$ fluxes.

Our analysis has also revealed the presence of a small number of
intermediate-age ($\sim 1$ Gyr-old), $\sim10^5$ M$_\odot$ clusters, as
well as up to a dozen old clusters resembling true globular clusters
(GCs), in relation to both their ages ($\sim 10$ Gyr) and masses ($\ga
10^5$ M$_\odot$). We emphasize that our discovery of a small number of
old GCs is secure. First, in our analysis we use a low metallicity of
$Z = 0.004 \, (0.2 {\rm Z}_\odot)$. It is expected that the oldest
stellar population components in a given galaxy are characterized by
the lowest metallicity, since they were formed at a time when there
was little recycled gas in the interstellar medium. (This argument is
valid under the assumption that these clusters formed in NGC 5253 and
were not subsequently accreted; see below.) Our age estimates at the
current low metallicity are already almost as old as the oldest ages
in our SSP models, whereas at such old ages SSP models are quite
insensitive to changes caused by metallicity differences. We confirmed
that adopting an even lower metallicity will not result in a different
conclusion.

In Fig. \ref{agemass.fig}, we have also indicated the expected
detection (50 per cent completeness) limit based on standard SSP
analysis (solid line); the limit is a steeply increasing function of
cluster mass with increasing age. Although we did not set out to
select a statistically complete sample of star clusters, it is
satisfying to note that our sample clusters obey a similar,
galaxy-wide average detection limit across the entire age range. We
focus specifically on resolved yet compact clusters (for which it is
straightforward to obtain the integrated photometry and a generic
completeness limit) rather than on more dispersed stellar associations
(cf. Annibali et al. 2011). This also enables a better comparison with
previous studies of this galaxy, many of which were based on
lower-resolution images than used in the present study.

The selection of the final sample shown in Fig. \ref{agemass.fig} is
governed by our observations with the lowest S/N. Depending on the
nature of the clusters (i.e., with or without an IR excess), the
images with the lowest S/N are the F330W and the F222M observations,
respectively. The complex behaviour of the selection limit resulting
from the data reduction and analysis steps, combined with the galaxy's
highly variable background, renders determination of the `true'
detection limit less than trivial. In addition, the effects of
stochasticity will also play a role in the (non-)detection of the
clusters closest to the detection limit (e.g., Silva-Villa \& Larsen
2010, 2011). Keeping in mind both of these issues, so that it is
unrealistic to assign a single detection limit to the entire galaxy,
the distribution of the data points in Fig. \ref{agemass.fig} implies
that the {\it average}, galaxy-wide detection limit is $M_V \simeq
-5.4$ mag ($V \simeq 22.1$ mag), but we note that local variations,
combined with the effects of stochasticity, could be up to 0.3 mag.

Given the number and concentration of old GCs above our detection
limit in the NGC 5253 data, we speculate that these are merely the
surviving high-mass, high-luminosity clusters of an initially much
larger population of clusters that formed around the time of galaxy
formation, many of which may have been disrupted or naturally faded to
below the detection limit because of stellar evolution. The young
clusters are distributed in two narrow `chimneys' near $\log(t \mbox{
  yr}^{-1}) = 6.7$ and 7.2, which are caused by the features in SSP
models around that age (cf. the `wiggles' in the SSP model used to
indicate the selection limit in Fig. \ref{agemass.fig}), but are not
necessarily associated with stochastic sampling of their mass
functions. In reality, we would expect these clusters to be spread out
more evenly across this young age range.

The clusters in both chimneys are distributed randomly throughout the
galaxy, similarly for objects with both $\log( t \mbox{ yr}^{-1}) \le
7.0$ and $7.0 < \log( t \mbox{ yr}^{-1}) \le 8.0$. The YSCs
characterized by a clear IR excess are confined to the galaxy's main
body; the intermediate-age clusters in NGC 5253, with ages $\log( t
\mbox{ yr}^{-1}) \simeq 9$, are found in the galaxy's outer region
(cf. Harbeck et al. 2012), towards the south.

Finally, the complement of $\ga 10$ Gyr-old clusters are predominantly
located in the southeast of the main body of NGC 5253
(cf. Fig. \ref{sources.fig}). Since our observational data reach
similar depths across the galaxy (except for the central region on the
East featuring a wide dust lane), we would have expected to detect
such old massive clusters across the galaxy with a similar level of
confidence. As such, we conclude that this spatial bias is likely
real. 

If these clusters were moving on chaotic, freely tumbling orbits
through the galaxy, one would have expected them to be well mixed on
these long timescales. However, Kobulnicky \& Skillman (1995) showed,
based on observed H{\sc i} gas dynamics, that the main body of the
galaxy appears to be rotating as a solid body along the galaxy's {\it
  major} axis. Under these circumstances, clusters that formed at one
end of the gaseous central bar or disc, may have remained confined to
a small area since the time of their formation. Alternatively, the old
clusters could have been accreted externally: to explain the galaxy's
peculiar velocity field, Kobulnicky \& Skillman (1995) also suggested
that accretion of a low-metallicity, gas-rich companion galaxy on a
higly inclined orbit is highly plausible (but see Davidge 2007). This
interpretation is supported by the unusual weakness of the galaxy's CO
emission, the morphology of the CO gas and its unusual kinematic
properties (Turner et al. 1997).

\subsection{Red- and infrared-excess sources}

We also note that almost all clusters affected by an
$\mbox{(infra-)}$red excess, are found among the YSC population in
clusters with masses $\la 10^4$ M$_\odot$. Adamo et al. (2010a,b,
2011a,b) identified two possible reasons for the occurrence of an
(infra-)red excess in more massive clusters in their sample BCG
galaxies. For clusters younger than approximately 6 Myr, they
suggested that the IR excess could be due to diffuse, hot dust in
which these YSCs may still be embedded and/or a large fraction of
young stellar objects and pre-main-sequence (PMS) stars. In the
optical $I$ band (F814W), the shallower red excess observed in such
YSCs may be owing to a dust photoluminescence bump in their SEDs at
0.7--0.9 $\mu$m, often seen in the presence of very strong UV
fields. Older clusters, with ages $\ga 10$ Myr, may exhibit an IR
excess, which these authors attribute to the uncertain treatment of
RSG stars in SSP models or the effect of stochasticity in the stellar
mass function. In NGC 5253, we have probed to much lower masses than
Adamo et al. (2010a,b, 2011a,b) could reach in their more distant BCG
sample.

Most of our IR-excess clusters are found in the younger `chimney',
characterized by an age of $\sim 5 \times 10^6$ yr. This is close to
the lower age limit of our SSP models and also close to the age when
YSCs are expected to become optically visible (i.e., after dispersion
of at least some of the embedded dust associated with their formation
process), so that a number of these clusters may have even younger
ages. Based on the detailed studies referenced above, these clusters
are most likely affected by a combination of stochastic sampling
effects and the presence of PMS stars. The IR-excess clusters in the
older chimney, at $\log(t \mbox{ yr}^{-1}) = 7.1$--7.2 are also of
relatively low mass and, hence, we expect them to be affected by
stochastic sampling effects, while at this age the stochastic colour
variations owing to RSG stars will also become apparent.

Only two of our IR-excess clusters allow us to offer a less ambiguous
explanation. The highest-mass cluster affected by an IR excess has a
mass of almost $10^5$ M$_\odot$. Although it has been assigned an age
of $\sim 4\times 10^6$ yr, stochastic sampling should only play a
minor role, so that we are confident that the excess in this case is
caused by the presence of a hot, diffuse dust cocoon or perhaps young
stellar objects. The oldest cluster affected by an IR excess has an
age of almost $10^8$ yr. Although it has a low mass of $\sim 6 \times
10^3$ M$_\odot$, so that stochastic sampling effects must duly be
taken into account, this is also the time when (relatively
short-lived) AGB stars appear in realistic stellar populations,
playing a similar role as RSG stars at ages near $10^7$ yr, both in
terms of the numbers present at a given time and their short
lifetimes.

\subsection{Comparison with previous studies based on fully sampled stellar mass functions}

Cresci et al. (2005) published the most recent statistical study of
the NGC 5253 cluster population (see also Harris et al. 2004). They
based their results on {\sl HST} F547M and F814W broad-band
observations and ground-based $K_{\rm s}$-band observations with the
Very Large Telescope, complemented with {\sl HST}-based H$\alpha$
data. Qualitatively, our results are similar to theirs, although there
are a number of important differences. They divide their final cluster
sample of 115 sources into a `young' subsample of 51 objects which
were detected in the H$\alpha$ filter, and an `old' subsample.

For the `young' subsample, they derive ages spanning the range from 3
to 19 Myr. This corresponds to the bulk of the clusters in our two
young chimneys in Fig. \ref{fig7.fig}. To derive the cluster masses,
they assumed an average extinction of $A_V = 0.98$ mag -- which
appears reasonable for most clusters based on our extinction estimates
in Fig. \ref{fig5.fig} and assuming a Calzetti et al. (2000)
extinction law -- ≈ßand an average age of 8 Myr; for the `old' clusters
they assumed an average age of 20 Myr (cf. Harris et al. 2004). Above
their 50 per cent completeness limit, they find a steeply declining
number of clusters with increasing mass, with a maximum mass of $\sim
10^6$ M$_\odot$. Our mass function (Fig. \ref{fig7.fig}) samples
significantly lower-mass clusters.

Contrary to their analysis, however, we believe that we do not have
sufficient numbers of clusters to comment on the reality (or
otherwise) of a `peak' in the mass function. Their result may be an
artefact of the steeply increasing incompleteness limit with
increasing age which could mask the true underlying mass distribution
and produce an artificial turnover in the observed mass distribution
(cf. Fig. \ref{agemass.fig}).

Prior to their work, Calzetti et al. (1997) used optical {\sl
  HST}/WFPC2 broad- and narrow-band observations in a careful attempt
to characterize the star cluster properties in the core of NGC
5253. Tremonti et al. (2001) used {\sl HST}/STIS UV spectroscopy of
four of the brightest clusters in the galaxy's central region to
derive their fundamental properties. Of the four clusters examined in
detail in both the Calzetti et al. (1997) and Tremonti et al. (2001)
studies, we managed to produce satisfactory results for three using
our broad-band-only SED fitting routines. Using Calzetti et al.'s
(1997) nomenclature, these are clusters NGC 5253-1, 2 and 6
(corresponding to, respectively, objects 129, 45 and 36 in Table
\ref{Yggdrasil.tab}). For NGC 5253-1, Calzetti et al. (1997) derive an
age of 8--12 Myr, while we find a best-fitting age of $14 \pm 3$
Myr. Calzetti et al. (1997) and Tremonti et al. (2001) find ages of,
respectively, 50--60 and $8^{+2.6}_{-0.9}$ Myr for NGC 5253-2,
compared with our best age estimate of $5.0 \pm 1.6$ Myr. Our results
thus support the younger age advocated by Tremonti et al. (2001),
although our mass estimate, $(6.4 \pm 1.3) \times 10^4$ M$_\odot$ is
somewhat higher than theirs, $\sim 1 \times 10^4$ M$_\odot$. Note that
although this cluster's derived mass tends towards the mass range
where the effects of stochastic sampling become smaller, we caution
that a small number of luminous red stars could easily affect its
broad-band SED to mimic a younger age. Finally, for NGC 5253-6,
Calzetti et al. (1997) find a best-fitting age of 10--17 Myr, which is
a close match to our age estimate, $14 \pm 2$ Myr.

We also compared our derived age and mass estimates with the
equivalent results of Harris et al. (2004). Within the uncertainties
associated with both sets of parameter determinations, our age and
mass determinations of the dozen objects with parameter determinations
in common are fully consistent with theirs (not shown).

Recently, Harbeck et al. (2012) reported the discovery of three
potentially massive clusters ($M_{\rm cl} \ge 10^5$ M$_\odot$) with
ages of order 1--2 Gyr in the galaxy's outer regions, based on {\sl
  HST}/ACS imaging observations in the F415W, F555W and F814W
filters. These clusters, which they refer to as \#2, 6 and 7, form
part of a more extensive population of intermediate-age and old
clusters. From an initial focus on the galaxy's starburst core and its
properties, more recent studies have attempted to explore the galaxy's
overall star-formation history. The intermediate-age clusters of
Harbeck et al. (2012), as well as our larger population of
intermediate-age and old clusters support the notion that NGC 5253 is
a very active dwarf starburst galaxy that has undergone multiple
episodes of star and star cluster formation (see also Calzetti et
al. 1997 and McQuinn et al. 2010a,b for similar conclusions based on
studies of the field-star population).

\subsection{The cluster size distribution in the context of previous work}
\label{sizes.sec}

Next, we consider the cluster size distribution shown in
Fig. \ref{sizes.fig} in the context of previous work. Harbeck et
al. (2012) tabulate the FWHMs of the 28 objects in the galaxy's outer
regions which they discuss (18 likely clusters and 10 background
galaxies). Their resulting size distribution (converted to Gaussian
$\sigma$'s) is generally flat, however, and does not show a clear peak
at smaller sizes similar to what we find in this paper. This is likely
caused by their cluster selection procedure, which was based on visual
inspection of the ACS images. In addition, we cannot directly compare
our cluster parameters derived in this paper with Harbeck et al.'s
(2012) results, because both samples are complementary, without any
spatial overlap.

Harris et al. (2004) also examined the NGC 5253 cluster sizes and
include a figure showing their (FWHM) size distribution as a function
of cluster magnitude. Their size distribution is based on analysis of
their F547M image (which is closest in wavelength to the F555W filter
we used). The vast majority of their sample clusters have sizes which
are similar to the PSF size, with a small tail towards more extended
clusters. Although this trend is qualitiatively similar to our result,
we point out that their size measurements were based on WFPC2/WF
camera images, which have a factor of approximately 2 lower resolution
than the ACS/HRC images upon which we base our analysis. The bottom
inset in Fig. \ref{sizes.fig} shows the equivalent figure for our data
set, also using the F547M magnitudes. The trend shown is similar to
that found by Harris et al. (2004). We re-emphasize here that we
specifically selected objects that appeared compact, which implies
that we selected against including stellar associations.

Finally, we consider the cluster size distribution in a wider
context. Bastian et al. (2005) examined the effective radius ($R_{\rm
  eff}$) distribution of their YSC sample in M51 and of the old GCs in
the Milky Way. They represented the distribution by a power law of the
form $N(R) {\rm d}r \propto r^{-\eta} {\rm d}r$, and found
best-fitting power-law indices of $\eta = 2.2 \pm 0.2$ and $2.4 \pm
0.5$ for the M51 and Milky Way clusters, respectively. This compares
to $\eta = 3.4$ for the YSCs in the merger remnant galaxy NGC 3256
(Ashman \& Zepf 2001). Here, we find a best-fitting power-law
distribution of the clusters' Gaussian $\sigma$'s characterized by
$\eta = 2.8 \pm 0.6$, which agrees to within the uncertainties with
these previous determinations -- provided that there is a one-to-one
correlation between $R_{\rm eff}$ and $\sigma_{\rm G}$, which is a
reasonable assumption if cluster profiles do not differ too
significantly among galaxies.

\section{Summary and conclusions}
\label{summary.sec}

Because of the wealth of available multi-wavelength data sets in the
{\sl HST} Data Archive, we chose NGC 5253 as target to compare two
model approaches to determine fundamental star cluster parameters
based on up-to-date input physics. We have gone significantly beyond
previous studies of the NGC 5253 star cluster population.

First, although we still used medium-, broad-band and H$\alpha$
imaging observations, we base our results on the highest-achievable
spatial resolution. The gain in resolution compared to previous
studies is of order a factor of two in both dimensions, while our
accessible wavelength range transcends previous studies by
incorporation of both the NUV and NIR passbands. Crucially, this
allows us to at least partially break the age--metallicity and
age--extinction degeneracies. Second, we applied two spectral
synthesis methods to our {\sl HST} observations in up to 10 filters,
taking into account the updated effects of nebular emission and the
improved physical understanding of the excess flux at red optical and
IR wavelengths found for a subset of young clusters.

We used the most recent models to explore the origin of excess
emission of a fraction of the YSCs in optical $I$ and
longer-wavelength NIR filters compared to `standard' SSP models,
assuming fully sampled stellar mass functions. Of the 182 objects in
our sample with reliable photometry in at least 7 filters (i.e.,
uncertainties $< 0.3$ mag), we obtained best fits of sufficient
quality (i.e., with sufficiently small $\chi^2_{\nu}$ and $Q > 0.001$)
for 149 cluster candidates; we discarded the remaining 33 objects from
our final sample. Of these 149 clusters, 30 were characterized by a
clear IR excess, while the remaining 119 objects exhibited SEDs as
expected for fully sampled clusters. We compared our derived cluster
ages and masses based on application of our novel Yggdrasil-based
approach (Adamo et al. 2010a,b, 2011a,b) with the results from
application of {\sc galev}+{\sc AnalySED} SED matching. The main
features in the overall age and mass distributions were well
reproduced by both independent approaches. Assuming a simple
underlying Gaussian distribution, the peak in the age distribution
resulting from our {\sc galev}+{\sc AnalySED} analysis occurs at
$\langle \log(t \mbox{ yr}^{-1}) \rangle = 7.11$, with
$\sigma_{\langle \log t \rangle} = 0.80$. For comparison, the
Yggdrasil-based analysis returned $\langle \log(t \mbox{ yr}^{-1})
\rangle = 7.10$, with $\sigma_{\langle \log t \rangle} = 0.94$. For
the mass distributions, we find $\langle \log(M_{\rm cl}/{\rm
  M}_\odot) \rangle = 3.38$ versus 3.37, and $\sigma_{\langle \log
  M_{\rm cl} \rangle} = 0.46$ versus 0.77, respectively.

The NGC 5253 cluster population is dominated by a significant number
of relatively low-mass ($M_{\rm cl} \ll 10^5$ M$_\odot$)
objects. Assuming fully sampled stellar mass functions, they have ages
ranging from a few $\times 10^6$ to a few $\times 10^7$ yr. This age
range is in excellent agreement with the starburst age of the host
galaxy ($\le 10$ Myr). Most of our IR-excess clusters are found in the
younger `chimney' observed in the diagnostic age--mass plane based on
our fully sampled SSP analysis, characterized by an age of $\sim 5
\times 10^6$ yr and masses of up to a few $\times 10^4$ M$_\odot$. The
observational properties of these clusters are most likely affected by
a combination of stochastic sampling effects and the presence of PMS
stars. The IR-excess clusters in the older chimney, at $\log(t \mbox{
  yr}^{-1}) = 7.1$--7.2 are also of relatively low mass and, hence, we
expect them to be affected by stochastic sampling effects, while at
this age the stochastic colour variations owing to luminous red stars
(including RSGs) will also become apparent.

Our analysis also revealed the presence of a small number of
intermediate-age ($\sim 1$ Gyr-old), $\sim10^5$ M$_\odot$ clusters, as
well as up to a dozen old clusters resembling true GCs, in relation to
both their ages ($\sim 10$ Gyr) and masses ($\ga 10^5$ M$_\odot$). Our
detection of this fairly small number of true GCs is simply owing to
selection effects and may not represent a genuine dearth of old,
low-mass clusters in the galaxy. The presence of populations of young,
intermediate-age and old clusters supports the notion that NGC 5253 is
a very active dwarf starburst galaxy that has undergone multiple
episodes of star cluster formation. The steeply increasing detection
limit as a function of increasing age implies that even if there were
large numbers of intermediate-age and old clusters of intermediate
masses, we would not be able to detect them at the present time.

\section*{Acknowledgments}

We thank Angela Adamo for providing the Yggdrasil-based analysis. We
also acknowledge her and Ralf Kotulla for their input in many of the
technical aspects of this work. We really appreciate the referee's
useful comments and constructive suggestions leading to the final
version of this manuscript. This paper is based on archival
observations with the NASA/ESA {\sl Hubble Space Telescope}, obtained
from the ST-ECF archive facility. This research has also made use of
NASA's Astrophysics Data System Abstract Service. RdG and PA
acknowledge research support from the National Natural Science
Foundation of China (NSFC) through grant 11073001. G\"O is a Royal
Swedish Academy of Sciences Research Fellow supported by a grant from
the Knut and Alice Wallenberg Foundation. EZ and G\"O acknowledge
financial support from the Swedish Research Council (VR) and the
Swedish National Space Board.

\end{document}